\documentclass[preprint,showpacs,preprintnumbers,amsmath,amssymb,prb]{revtex4}
\usepackage{graphicx}
\usepackage{dcolumn}
\usepackage{bm}
\usepackage{epsfig}
\usepackage{subfigure}

\begin{document}

\preprint{PREPRINT}

\title{Thermodynamic, Dynamic and Structural Anomalies for Shoulder-like potentials}

\author{Ney M. Barraz Jr.}
\affiliation{Instituto de F\'{\i}sica, Universidade Federal do Rio Grande do
Sul, Caixa Postal 15051, 91501-970, Porto Alegre, RS, Brazil}

\author{Evy Salcedo}

\affiliation{Instituto de F\'{\i}sica, Universidade Federal do Rio Grande do
Sul, Caixa Postal 15051, 91501-970, Porto Alegre, RS, Brazil}

\author{Marcia C. Barbosa}
\affiliation{Instituto de F\'{i}sica, Universidade Federal do Rio Grande do Sul\\
Caixa Postal 15051, 91501-970, Porto Alegre, Rio Grande do Sul}

\date{\today}
\begin{abstract}
Using molecular dynamic simulations we study a family of continuous
core-softened potentials consisting of a hard core, a  shoulder
 at closest distances and  an attractive  well at further distance.
The repulsive shoulder and the well distances represent two length scales.
We show that if the first scale, the shoulder, is repulsive or
has a small well, the potential has a region in the 
pressure-temperature phase diagram with density, 
diffusion and structural  anomalies. However, if the closest scale
becomes a deep attractive well the regions in the pressure-temperature
phase diagram where the three anomalies are present shrink and 
disappear. This result enables us to predict by
the shape of the core-softened  potential if anomalies would or would not
be present.

\end{abstract}
\maketitle

\section{\label{sec1}Introduction}

Most liquids contract upon cooling. This is not the case of water, a liquid
where the specific volume at ambient pressure starts to increase 
when cooled below $T\approx4 ^oC$~\cite{Wa64}\cite{An76}. Besides, in a certain
range of pressures, also exhibits an anomalous increase of compressibility 
and specific heat upon cooling~\cite{Pr87,Pr88,Ha84}.
Experiments for Te, \cite{Th76} Ga, Bi,~\cite{LosAlamos} S,~\cite{Sa67,Ke83} 
and Ge$_{15}$Te$_{85}$,~\cite{Ts91}  and simulations for silica,~\cite{An00,Ru06b,Sh02,Po97}
silicon~\cite{Sa03} and BeF$_2$,~\cite{An00} show, as well, the same density anomaly.

Water also has dynamic anomalies. Experiments  show that the diffusion 
constant, $D$, increases on compression at low temperature, $T$,  up to a maximum $D_{\rm max} (T)$ at $p =
p_{D\mathrm{max}}(T)$. The behavior of normal liquids, with $D$  decreasing on compression, is 
restored in water only at high $p$, e.g. for $p > p_{D\mathrm{max}}\approx 1.1$ kbar  at $10^o$C ~\cite{An76,Pr87}
Numerical simulations for SPC/E water~\cite{spce} recover the experimental results and 
show that the anomalous behavior of $D$ extends to the metastable liquid phase of water at negative pressure,
a region that is difficult to access for experiments.~\cite{Ne01,Er01,Mi06a,Ku06} In this region the diffusivity $D$ decreases for decreasing $p$ until it reaches a minimum value $D_{\rm min} (T)$ at some pressure $p_{D\mathrm{min}}(T)$, and the normal behavior, with $D$ increasing for
decreasing $p$, is reestablished only for $p < p_{D\mathrm{min}}(T)$~\cite{Ne01,Er01,Mi06a,Mu05}. Besides water, silica~\cite{Sh02, Ch06} and silicon~\cite{Mo05} also exhibit a diffusion anomalous region.

It was proposed a few years ago that these anomalies are related to a second critical
point between two liquid phases, a low density liquid (LDL) and a high density liquid (HDL)~\cite{Po92}. This critical point was discovered by computer simulations. This work   suggests that this
critical point is located at the supercooled region beyond  the line of homogeneous nucleation and thus cannot be experimentally measured. Even with this limitation, this hypothesis has been supported by indirect experimental 
results~\cite{Mi98, Sp76}.

In order to describe the anomalies present in water and in other liquids, isotropic models has been used 
as  the simplest framework to understand the physics of the liquid-liquid phase transition and liquid state anomalies. From the desire of constructing a simple two-body isotropic potential capable of describing
the complicated  behavior present in water-like molecules, a number of models in which
single component systems of particles interact via core-softened potentials~\cite{De03} have been proposed. 
They possess a repulsive core that exhibits a region of softening where the slope changes dramatically. This region can be a shoulder or a ramp~\cite{Sc00,Fr01,Bu02,Bu03,Sk04,Fr02,Ba04,
Ol05,He05a,He05b,He70,Ja98,Wi02,Ma04,Ku04,Xu05,Ol06a,Ol06b,Ol07,Ol08a,Ol08b,Ol09}.
These models exhibit density, diffusion and structural anomalies,
but depending on the specific shape of the potential, the anomalies might be hidden in the metastable and unstable phases
\cite{Ol09}. The relation between the specific shape of the core-softened potential and 
the presence or not of the anomalies is still missing.

How the specific shape of a core-softened potential affects the location of the anomalies and the critical points? 
In order to answer to this question in this paper we analyze a family of continuous core-softened 
potentials that exhibit two
length scales, a shoulder followed by an attractive well. When the shoulder is purely repulsive, this 
core-softened potential represents  the effective pair interaction between two neighbors tetramers~\cite{Ol06a, Kr08}
and  the density, the diffusion and the structural anomalies are present~\cite{Ol06a}\cite{Ol06b}.
If the shoulder has a deep attractive well, this potential
 it is related to the effective interaction potential between two water molecules 
obtained from the ST4~\cite{He93} or  TIP5P~\cite{Ya08} models for water. In this case the effective potential is 
derived from the oxygen-oxygen  radial distributions function, solving the Ornstein-Zernike equation by using
an integral equation method~\cite{He93, Ya08}. The resulting potential has  a shoulder with a deep
 attractive well at closest distance and a second attractive well with lower energy at furthest distance. The detailed
depth of the
softening region depends on the approximations employed.
This potential leads, as we are going see in this paper,  to systems in which the anomalies 
are in the unstable region of the phase diagram while in the full ST4 and TIP5P systems the anomalies can be observed.
It is important, therefore, to understand what is lost when one goes from the specific anisotropic
ST4 and TIP5P potentials to the isotropic spherical symmetric case.

So, in this paper we study what happens with the region in the pressure-temperature phase diagram where
the anomalies are located as the potential changes from a repulsive shoulder to a very deep well. Our results will shade some light not only in the use of spherical symmetric approximations of asymmetric potentials but also will help to design potentials for new systems with anomalies.

The paper is organized as follows. In sec. II the family of potentials is introduced and its link with the derivation
the framework of the integral equations is presented. In sec. III these potentials are tested for presence density, diffusion and structural anomalies, and for the the existence 
of two liquid phases and a critical points by  molecular dynamic simulations. Conclusions are presented in sec. IV.

\section{\label{sec2} The Model}

We study a system of $N$ particles, with diameter $\sigma$, where the pair interaction
is described by a family of continuous potentials given by

\begin{equation}
  U(r) = \epsilon \left[ \left( \frac{\sigma}{r} \right)^{a} -
    \left( \frac{\sigma}{r} \right)^{b} \right] + \sum_{j=1}^{4}h_{j}
  \exp \left[ -\left( \frac{r-c_{j}}{w_{j}} \right)^{2} \right] \;\;.
  \label{eq:potential}
\end{equation}

\begin{figure}[h]
  \begin{centering}
    \includegraphics[width=8cm]{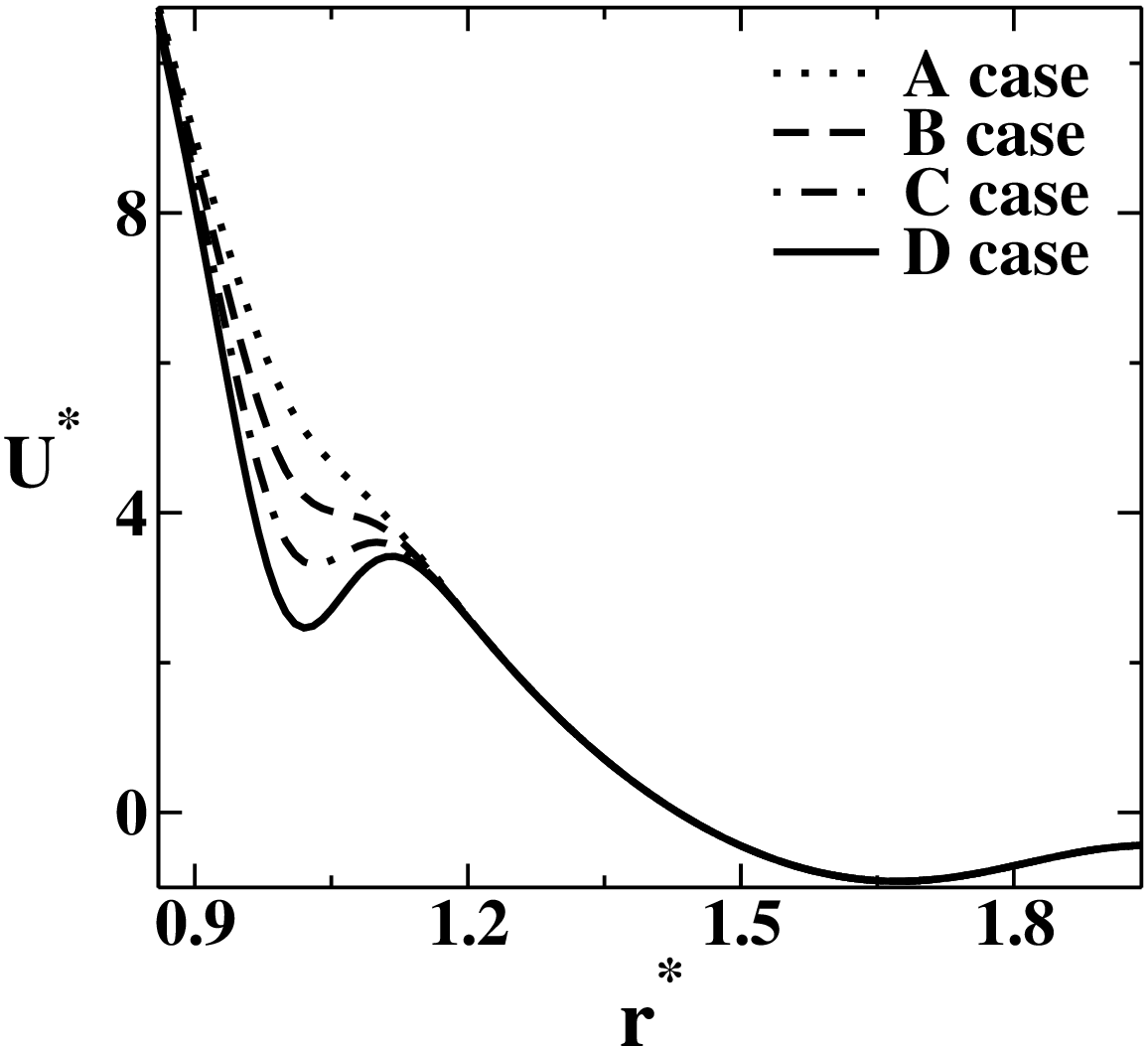} 
    \par
  \end{centering}
  \caption{Interaction potential obtained by changing parameters $h_1$ in
    Eq.~(\ref{eq:potential}). }
  \label{fig:potential} 
\end{figure}

The first term is a Lennard-Jones potential-like  and the second one is composed by four Gaussian, each
one centered in $c_j$. This potential can represent a whole family of 
intermolecular interactions, depending of the choice of the parameters $a,b,\sigma,\{h_j, c_j, w_j\}$, with $j=1, \dots, 4$. The parameters are chosen in order to obtain a two length scale potential \cite{He93}.

Modifying $h_1$ in the Eq.~\ref{eq:potential} allow us to change the depth of the hard-cor
e well, as illustrated in Fig.~\ref{fig:potential}. Here we use four different values for $h_1$ and they are expressed as a multiple of a reference value $h_1^{ref}$ as shown in the Table~\ref{table1}. For all the four cases the values of $a,b,\{c_j, w_j\}$ with $j=1, \dots, 4$ and $h^{ref}$ are given in the Table~\ref{table2}. The depth of the region
of softening  of  the potentials illustrated in the Fig.~\ref{fig:potential} where chosen so that the potential B is the shallow shoulder-like potential similar to the one  studied by de Oliveira et al.~\cite{Ol06a} that exhibits the anomalies, while
for the potential $D$ the region of softening has the same depth as the potential obtained by using the oxygen-oxygen 
radial distribution function for the ST4 model~\cite{He93}. For comparison we also analyzed two other cases: potential
$A$ with a ramp-like shoulder and potential $C$, with a very shallow shoulder.

The properties of the system were obtained by $NVT$ molecular dynamics using Nose-Hoover 
heat-bath with coupling parameter $Q = 2$. The system  is characterized by 500 particles 
in a cubic box with periodic boundary conditions, 
interacting with the intermolecular potential described above. All physical quantities 
are expressed in reduced units and defined as
\begin{eqnarray}
t^* &=& \frac{t(m/\gamma \epsilon)^{1/2}}{\sigma}\nonumber  \\
T^* &=& \frac{k_{B}T}{\gamma \epsilon}\nonumber  \\
p^*& =& \frac{p\sigma}{\gamma \epsilon}\nonumber \\
\rho^{*} &=& \rho \sigma^3 \nonumber \\
D^* &=& \frac{D m}{\epsilon \gamma \sigma^2}\nonumber
\end{eqnarray}
where $\gamma = 50$. Standard periodic boundary conditions together with predictor-corrector 
algorithm were used to integrate the equations of motion with a 
time step $\Delta t^{*}=0.002$ and potential cut off radius $r_{c}^{*}=3.5$.
The initial configuration is set on solid or liquid state and, in both cases, the
 equilibrium state was reached after $t_{eq}^{*}=1000$. From this time on the physical 
quantities were stored in intervals of $\Delta t_R^* = 1$ during
$t_R^* = 1000$. The system is uncorrelated after $t_d^*= 10$, from the velocity auto-correlation function. $50$
descorrelated samples were used to get the average of the physical quantities. The thermodynamic 
stability of the system was checked analyzing the dependence of pressure on density, by 
the behavior of the energy and also by visual analysis of the final structure, searching for cavitation.

\begin{center}
\begin{table}
\caption{Parameters $h_1$ for potentials A, B, C and D.}
\centering{}
\begin{tabular}{|c|c|}
\hline 
Potential &   Value   of $h_1$  \tabularnewline \hline \hline
$A$       & $ 0.25\, h_1^{ref}$  \tabularnewline \hline
$B$       & $ 0.50\, h_1^{ref}$ \tabularnewline \hline
$C$       & $ 0.75\, h_1^{ref}$ \tabularnewline \hline
$D$       & $ 1.00\, h_1^{ref}$        \tabularnewline \hline

\end{tabular}
\label{table1}
\end{table}
\end{center}
\begin{center}
\begin{table}
\caption{Parameters for potentials A, B, C and D.}
\centering{}
\begin{tabular}{|c|c|c|c|}
\hline 
Parameter &    Value     & Parameter & Value                       \tabularnewline \hline \hline
$a$       & \ $9.056$ \  & $w_{1}$   & \hspace{0.2cm} $0.253$      \tabularnewline \hline
$b$       & $4.044$      & $w_{2}$   & \hspace{0.2cm} $1.767$      \tabularnewline \hline
$c$       & $0.006$      & $w_{3}$   & \hspace{0.2cm} $2.363$      \tabularnewline \hline
$d$       & $4.218$      & $w_{4}$   & \hspace{0.2cm} $0.614$      \tabularnewline \hline
$c_{1}$   & $2.849$      & $h_{1}^{ref}$ & $-1.137$                \tabularnewline \hline
$c_{2}$   & $1.514$      & $h_{2}$   & \hspace{0.2cm} $3.626$      \tabularnewline \hline
$c_{3}$   & $4.569$      & $h_{3}$   &  $-0.451$                   \tabularnewline  \hline
$c_{4}$   & $5.518$      & $h_{4}$   & \hspace{0.2cm} $0.230$      \tabularnewline \hline
\end{tabular}
\label{table2}
\end{table}
\end{center}

\section{\label{sec3} Results}

\subsection*{Pressure-Temperature Phase Diagram}
\begin{figure}[h]
  \begin{centering}
    \begin{tabular}{cc}
      \includegraphics[clip=true,width=6cm]{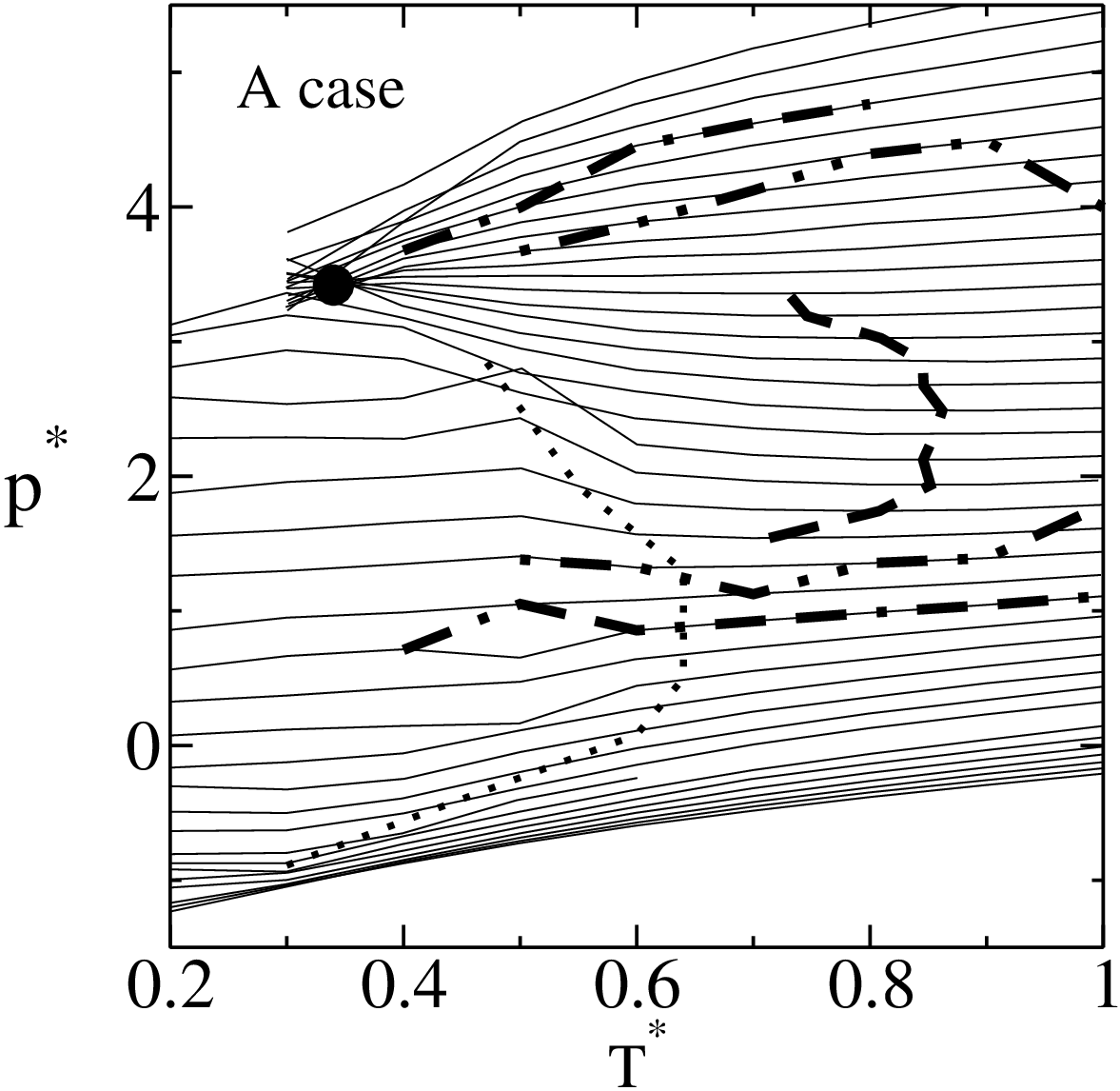} &
      \includegraphics[clip=true,width=6cm]{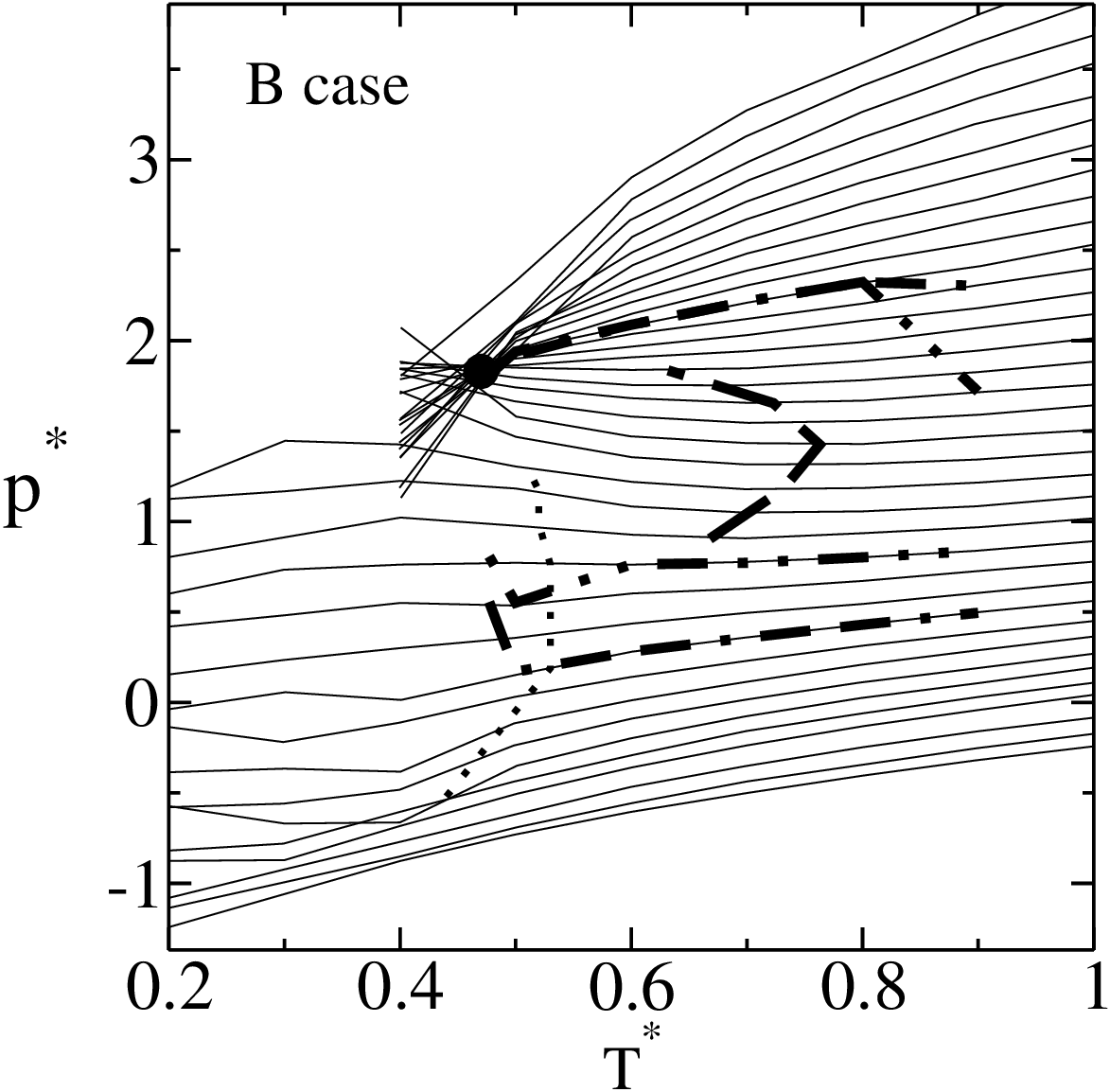} \tabularnewline
      \includegraphics[clip=true,width=6cm]{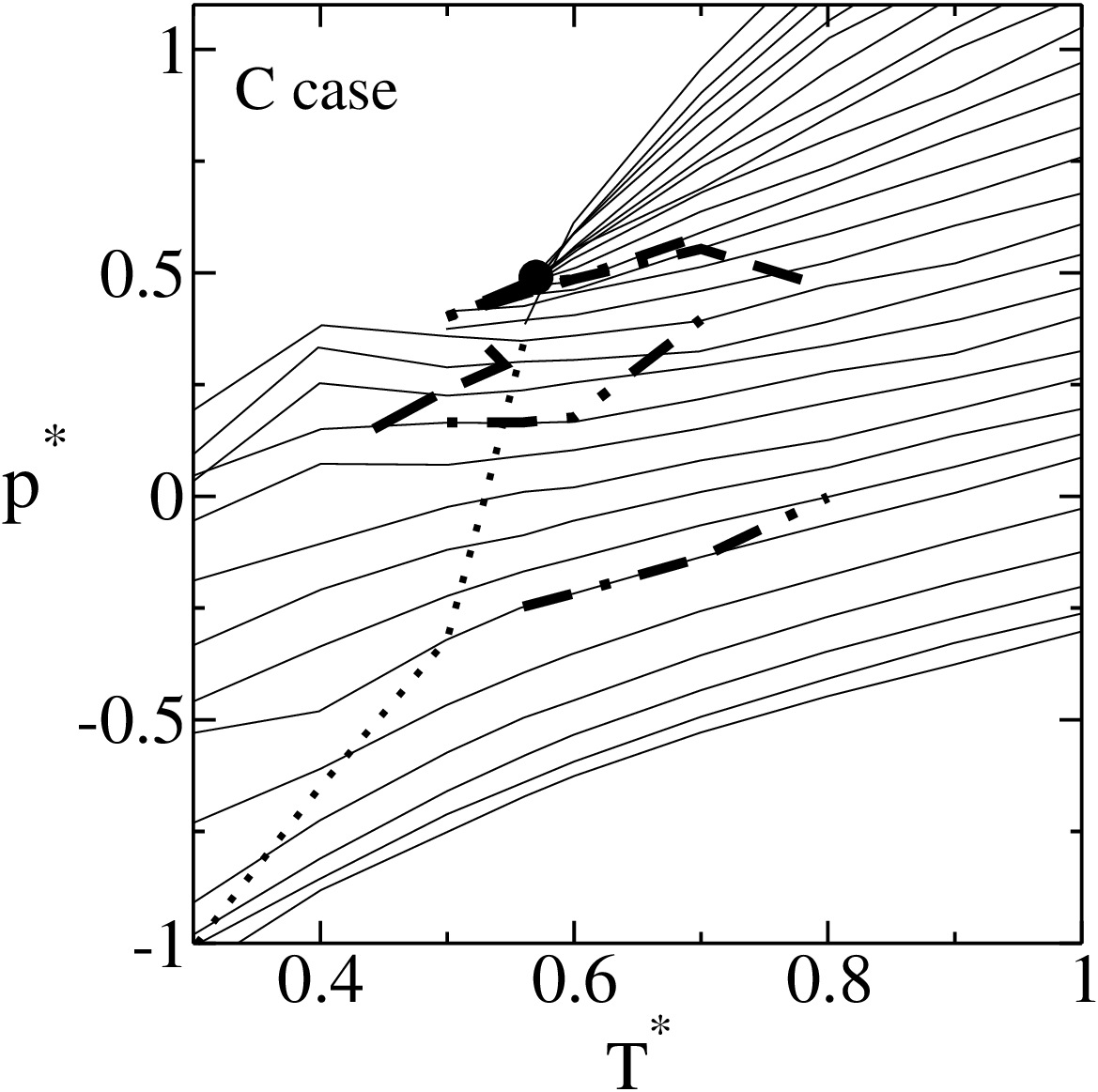} &
      \includegraphics[clip=true,width=6cm]{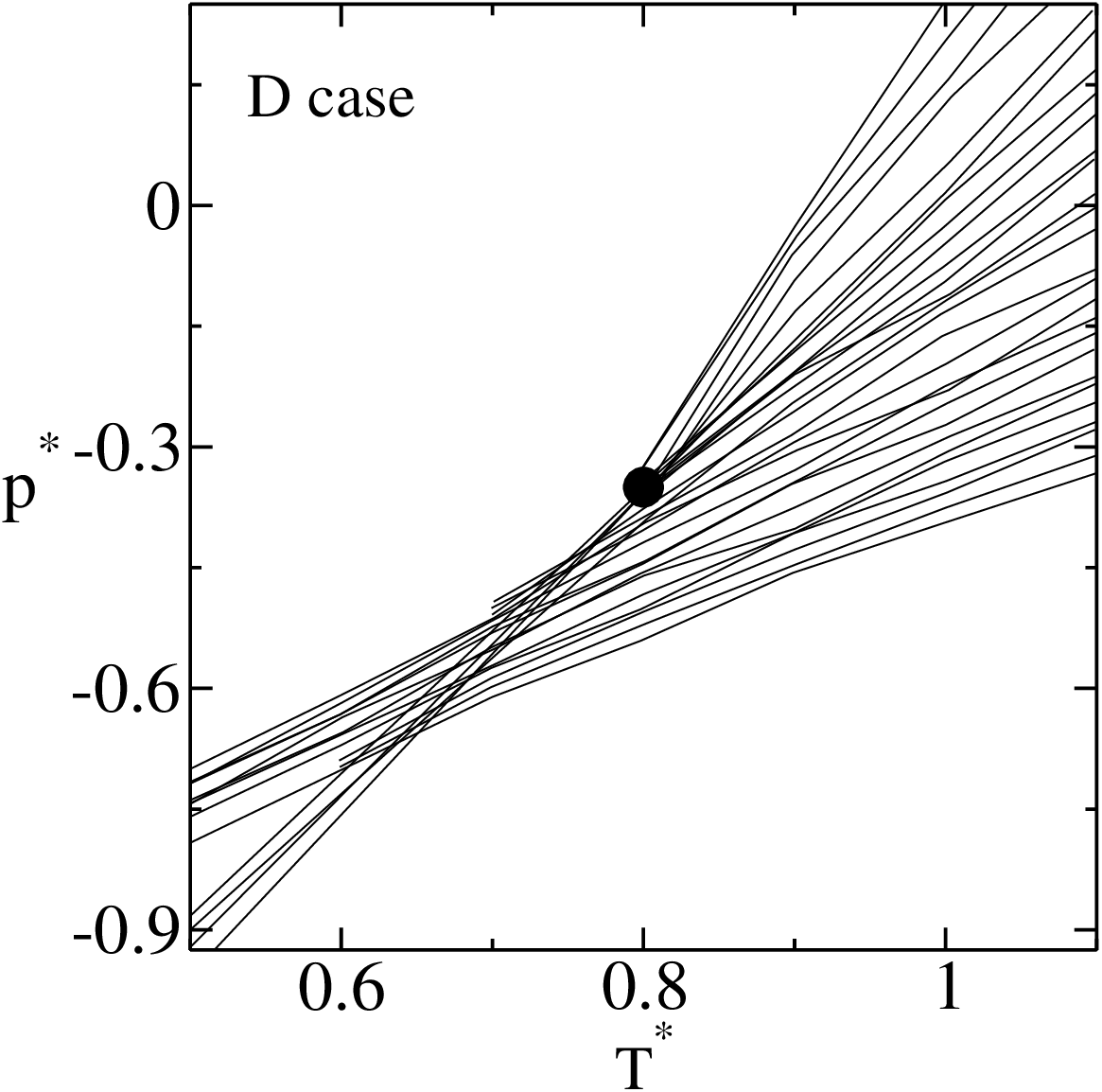} \tabularnewline
    \end{tabular}
    \par
  \end{centering}
\caption{Pressure-temperature phase diagram  for cases $A$, $B$, $C$ and $D$. The thin solid
lines are the isochores $0.30<\rho^*<0.65$. The liquid-liquid critical point is shown as a dot, the temperature of maximum density is a solid thick line, the diffusion extrema is the dashed line and the structural extrema is the dashed-dotted line. The dotted line indicates the limit between the fluid and the amorphous regions.}
\label{fig:PT} 
\end{figure}

First,  we are going to show the effects of the shoulder depth  in the presence or not of the thermodynamic anomalies and the location in the pressure-temperature phase diagram of the different phases.Fig.~\ref{fig:PT} illustrates the pressure-temperature phase diagram of the four studied cases. The system has at high temperatures a fluid phase and a gas phase (not shown). These two phases coexist at a first order line that ends at a critical point (see Table~\ref{table3} for the pressure and the temperature values). At low temperatures and high pressures there are two liquid phases coexisting at a first order line ending at a second critical point (see Table~\ref{table4} for  the pressure and the temperature values) that is identified in the graph by the region  where isochores cross. 

In the Fig.~\ref{fig:PT} at low temperatures and low
pressures the dotted line separates the fluid phase from the amorphous region where the diffusion 
becomes zero. For the potential $A$, the amorphous region is located in a pressure 
range $-0.61 \gtrsim p^* \gtrsim 3.40$, for $B$ case this region is located in the range $-0.75 \gtrsim p^* \gtrsim 0.40$ and for $C$ case it is located 
in the range $-0.75 \gtrsim p* \gtrsim 0.40$. The potential $D$  does not has a stable amorphous phase. Hence, 
as the shoulder becomes deeper the amorphous phase shrinks and moves to a lower pressure range. 

At low temperatures and high pressures two
liquid phases are present. As the shoulder becomes deeper the 
liquid-liquid coexistence line slides down to lower pressures and
it goes to higher temperatures. This indicates that the 
deeper the shoulder the liquid-liquid phase transition
stays stable for higher temperatures. Therefore, even thought this transition only exists if the
 attractive part of the potential is present (the second length scale), the stability of the 
liquid phases is determined by the depth of the shoulder (the first length scale).

\begin{center}
\begin{table}
\caption{Critical point location for potentials A, B, C and D.}

\centering{}
\begin{tabular}{|c|c|c|}
\hline 
Potential & $T_{c1}^*$  &  $p_{c1}^*$       \tabularnewline \hline \hline
$A$       & \ $1.93$ \    & \ $0.072$ \     \tabularnewline \hline
$B$       &   $1.98$      &   $0.078$       \tabularnewline \hline
$C$       &   $2.02$      &   $0.080$       \tabularnewline \hline
$D$       &   $2.15$      &   $0.094$       \tabularnewline \hline

\end{tabular}
\label{table3}
\end{table}
\end{center}

\begin{center}
\begin{table}
\caption{Second critical point location for potentials A, B, C and D.}

\centering{}
\begin{tabular}{|c|c|c|}
\hline 
Potential &   $T_{c2}^*$    &              $p_{c2}^*$     \tabularnewline \hline \hline
$A$       &  \ $0.35$ \     & \hspace{0.3cm} $3.44$ \     \tabularnewline \hline
$B$       &    $0.48$       & \hspace{0.2cm} $1.86$       \tabularnewline \hline
$C$       &    $0.57$       & \hspace{0.2cm} $0.49$       \tabularnewline \hline
$D$       &    $0.81$       &                $-0.33$      \tabularnewline \hline

\end{tabular}
\label{table4}
\end{table}
\par\end{center}

\begin{figure}[h]
  \begin{centering}
    \includegraphics[width=8cm]{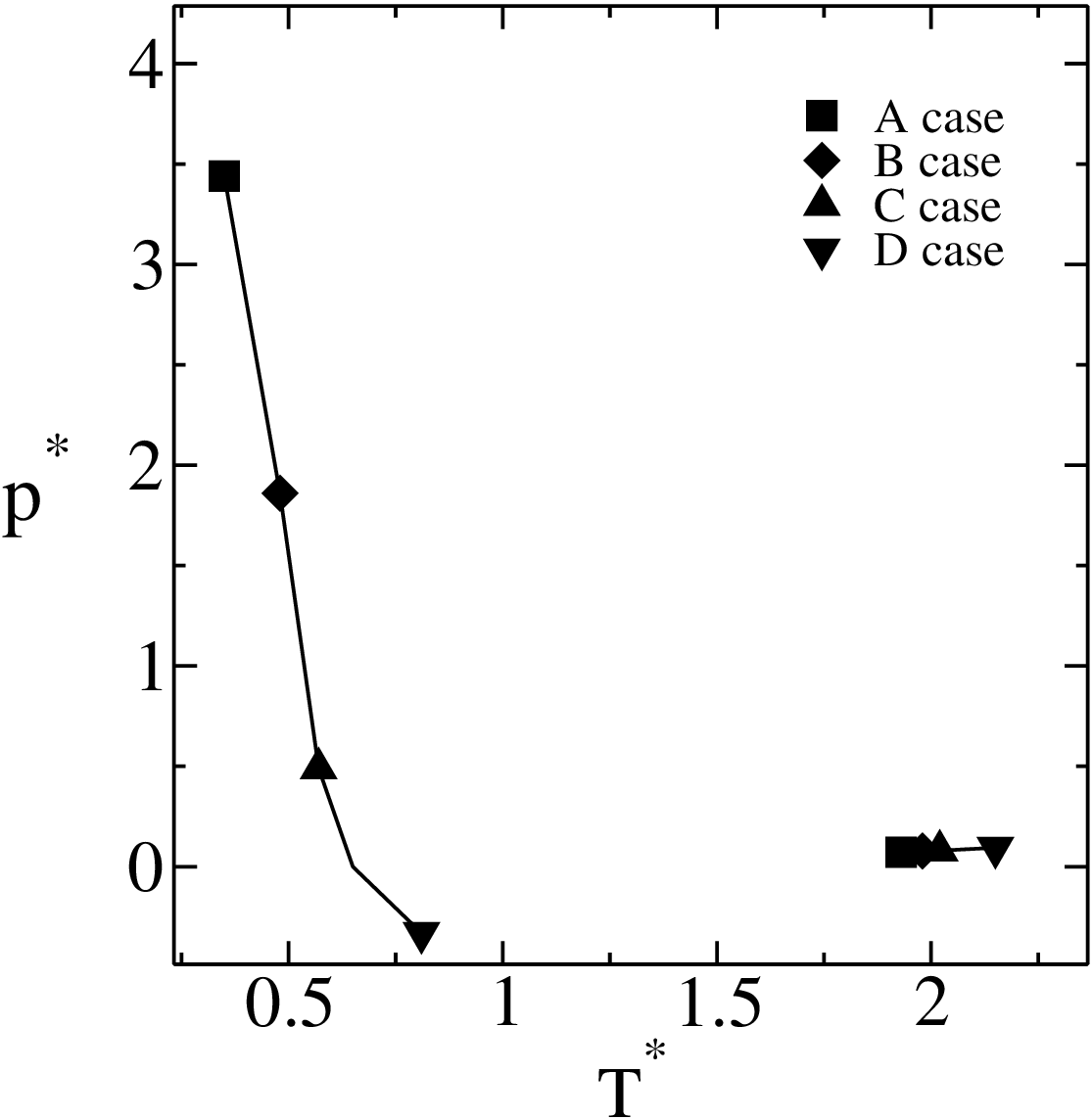} 
    \par
  \end{centering}
  \caption{Location of the critical points on pressure-temperature phase diagram for cases $A$, $B$, $C$ and $D$. }
  \label{fig:PT_CP} 
\end{figure}

\subsection*{Thermodynamics anomaly}

\begin{table}
  \begin{centering}
    \begin{tabular}{|cc|c|c|c|}
\hline 
cases &          & $p_l$  & $p_m$  & $p_h$   \tabularnewline \hline\hline
      & $\rho^*$ & $0.47$ & $0.52$ & $0.57$  \tabularnewline 
 $A$  & $T^*$    & $0.71$ & $0.85$ & $0.73$  \tabularnewline 
      & $p^*$    & $1.50$ & $2.50$ & $3.30$   \tabularnewline \hline\hline

      & $\rho^*$ & $0.46$ & $0.50$ & $0.54$  \tabularnewline
 $B$  & $T^*$    & $0.67$ & $0.76$ & $0.63$  \tabularnewline
      & $p^*$    & $0.90$ & $1.40$ & $1.80$   \tabularnewline \hline\hline

      & $\rho^*$ & $0.40$ & $0.42$ & $0.43$  \tabularnewline
 $C$  & $T^*$    & $0.44$ & $0.54$ & $0.52$  \tabularnewline
      & $p^*$    & $0.15$ & $0.29$ & $0.36$  \tabularnewline\hline
    \end{tabular}
    \par
  \end{centering}
  \caption{Limits values for density ($\rho^*$), temperature ($T^*$) and pressure ($p^*$) of the thermodynamics
anomalies on pressure-temperature diagram. Where $p_l$ is a lower limit (lesser pressure), $p_m$ is a inflection point of the anomaly (higher temperature) and $p_h$ is a higher limit of the anomalies (higher pressure).
    \label{TMD-tabla}}
\end{table}

The  Fig.~\ref{fig:PT} also shows the isochores $0.30 \leq \rho^*\leq 0.65$ 
represented by thin solid lines. The temperature of  maximum density 
at constant pressure coincides with the minimum pressure 
on isochores, $\left(\frac{\partial p}{\partial T}\right)_{\rho}=0$.
From the equation
\begin{equation}
  \left( \frac{\partial V}{\partial T} \right)_{p} = -\left(
  \frac{\partial p}{\partial T} \right)_{V} \left( \frac{\partial
    V}{\partial p} \right)_{T}
\label{tmd}
\end{equation}
is possible to see that, for a fixed density, a minimum in the pressure as a function of temperature 
represents a maximum in the density as a function of temperature, named temperature of maximum density (TMD)
given by
$\left( \frac{\partial V}{\partial T} \right)_{p} = 0$. The TMD  is the boundary of the region of thermodynamic 
anomaly, where a decrease in the temperature at constant
pressure implies an anomalous increase in the density and therefore an anomalous behavior of density (similar to what happens in water). Fig.~\ref{fig:PT} shows the TMD as a  
thick solid line. For the potentials $A$, $B$ and $C$ the TMD is present but for potential D no 
TMD is observed.

Similarly to what happens with the location of amorphous region and of the second critical, as the 
shoulder becomes deeper, the region in the pressure-temperature phase diagram delimited by the TMD goes to lower pressures, shrinks and disappears for the case $D$, the potential with the deepest shoulder. As the region delimited by the TMD shrinks, it also goes to lower temperatures. For the potential $C$ the temperature range of the TMD is lower than the liquid-liquid  critical point. The thermodynamic parameters that limits the TMD in phase diagram are shown in Table~\ref{TMD-tabla}, where $p_l$ represents the pressure of the  point with the lowest pressure, $p_h$ the pressure of the point with highest pressure and $p_m$ the pressure of the point with the highest temperature.

\begin{figure}[h]
  \begin{centering}
    \begin{tabular}{cc} \includegraphics[width=8cm]{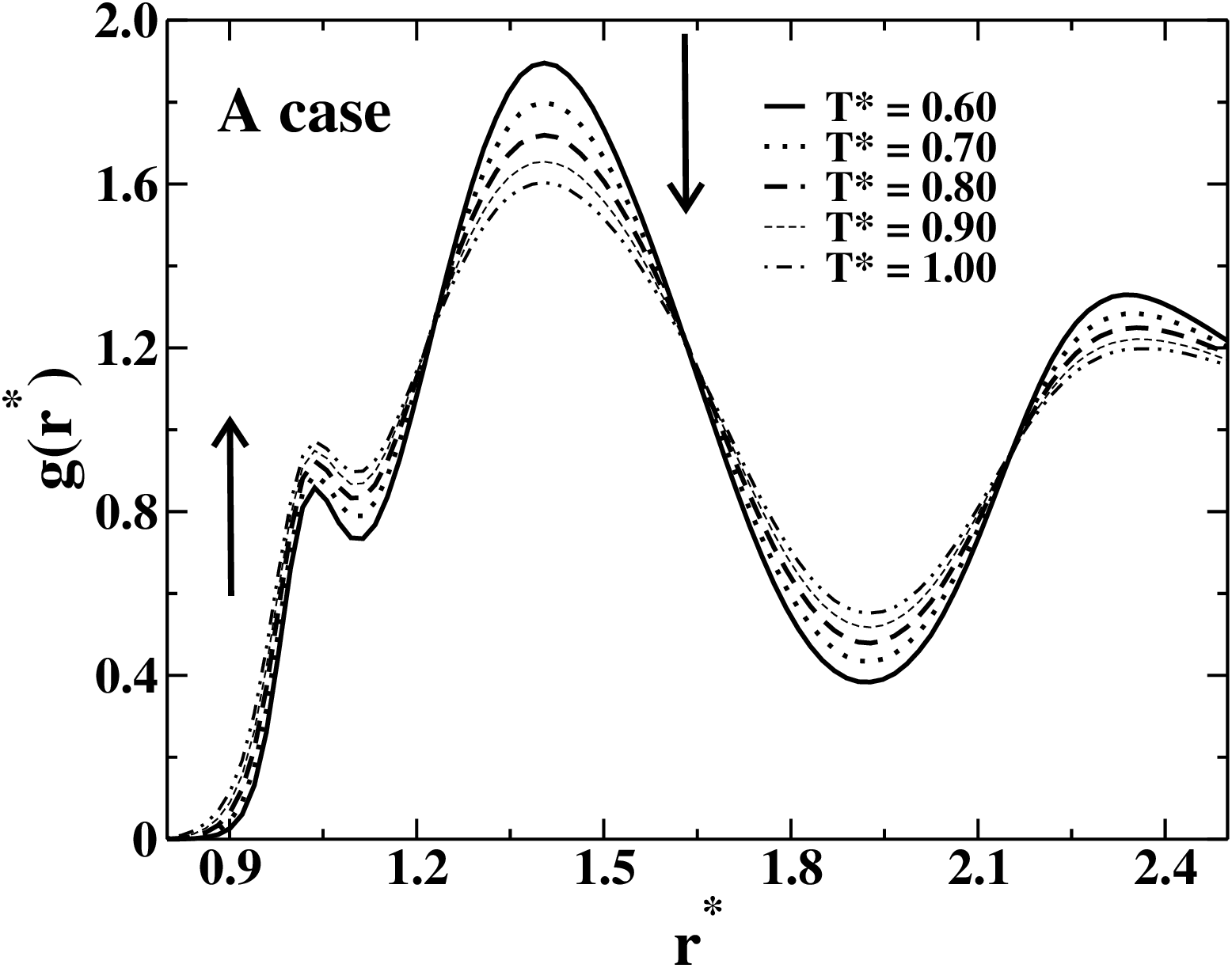} & 
      \includegraphics[width=8cm]{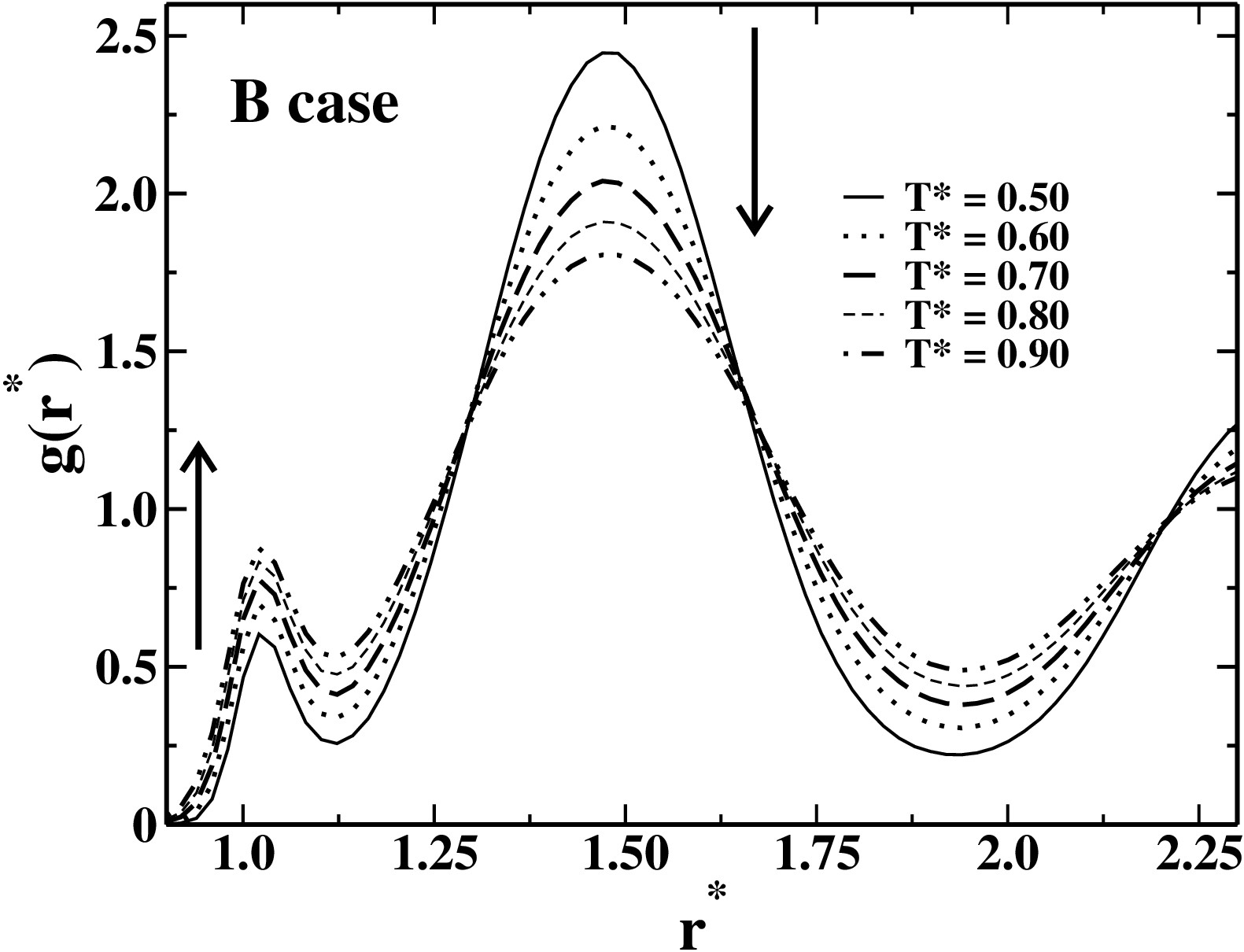} \tabularnewline \tabularnewline
      \includegraphics[width=8cm]{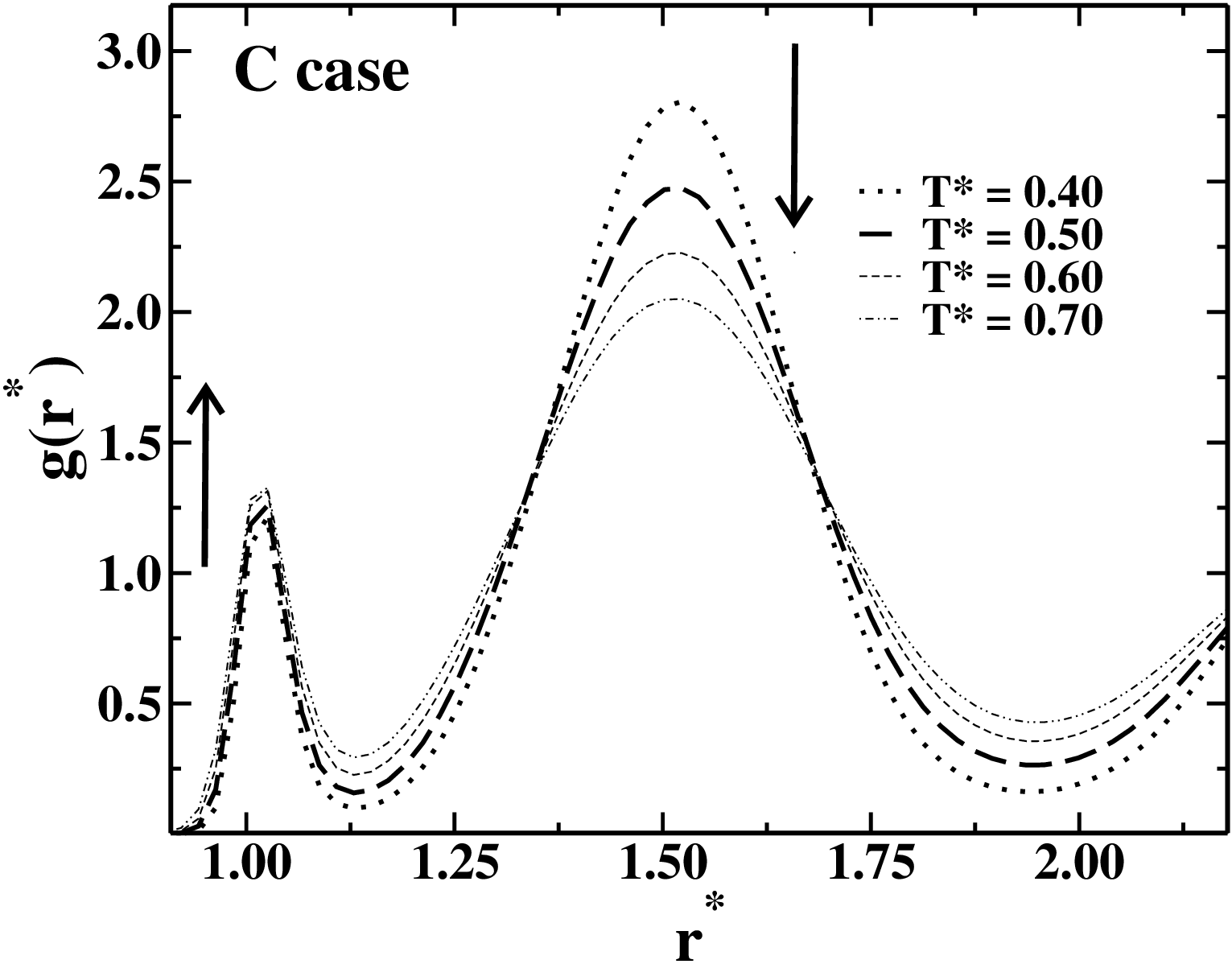} &
      \includegraphics[width=8.5cm]{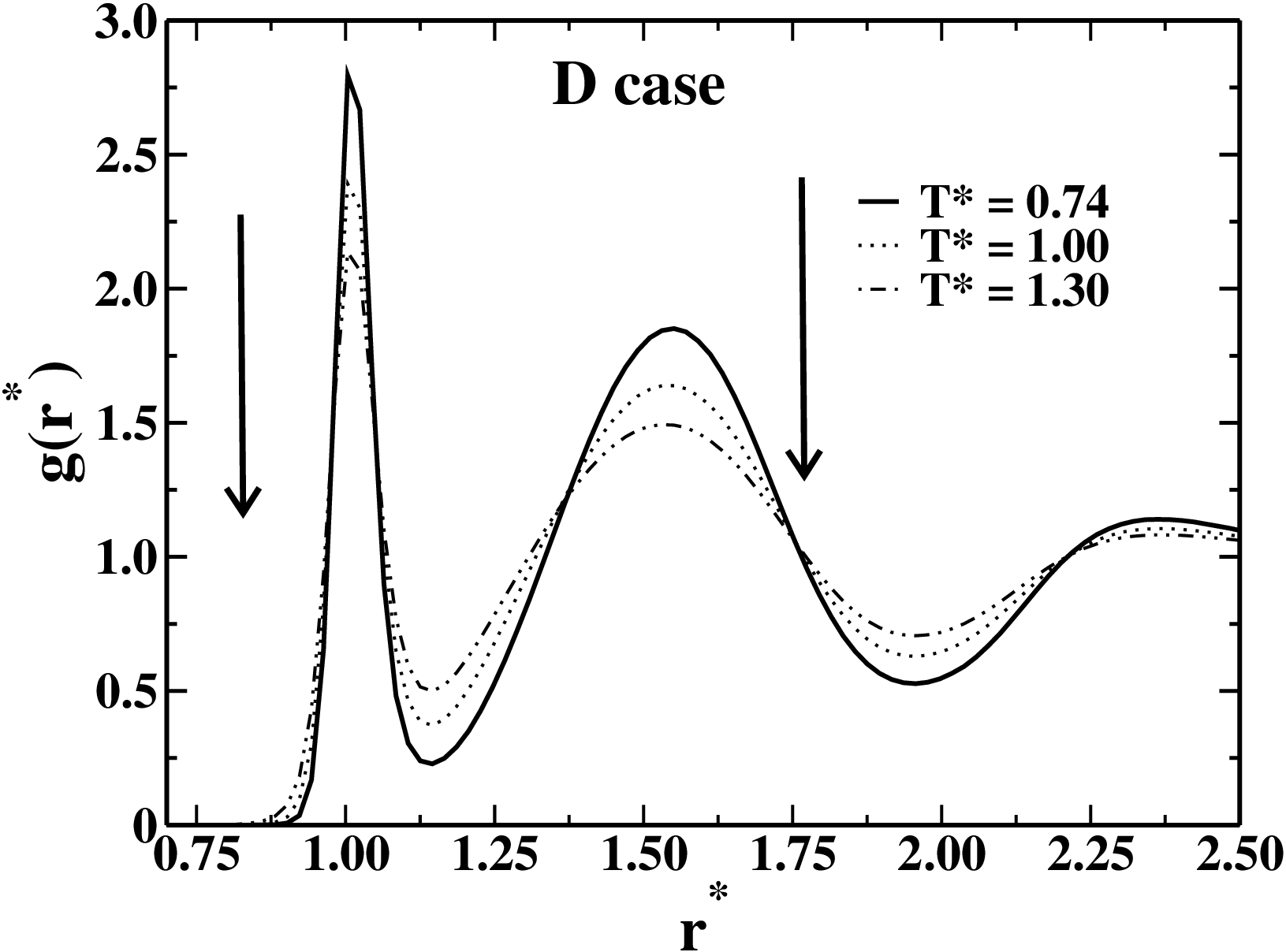} \tabularnewline\

    \end{tabular}
    \par
  \end{centering}
  \caption{Radius distribution as a function of the distance  for the four potentials. In the 
cases $A$, $B$ and $C$ the first peak of g$(r^*)$ grows in increasing temperature, while the second peak decreases. For the potential $D$ all the peaks decreases with temperature.}
  \label{fig:gr} 
\end{figure}

The link between the depth of the shoulder and the presence or 
not of the TMD goes as follows. The TMD is related to
the presence of large regions in the system in which particles are 
in two preferential distances represented by the first scale and the 
second scale in our 
potential ~\cite{St98,St99,St00,Ol09}.  While 
for normal liquids as the temperature is increased
the percentage of particles at closest scales 
decreases (see case $D$ in the Fig.~\ref{fig:gr}),
for the anomalous liquid (see cases $A$, $B$ and $C$ in the 
Fig.~\ref{fig:gr}) there are a region 
in the pressure-temperature phase diagram where 
as the  temperature is increased the percentage 
of particles at the closest distance increases. Particles
move from the second to the first scale. In the 
first case, the decrease of particles in the first scale
leads to a decrease of density with increase of temperature, behavior
expected for normal liquids. In the second case, the increase of
particles in the first scale leads to an increase of density
with temperature what characterizes the anomalous region.
Notice that as the temperature is increased particles move
from the second scale to the first scale.
The anomaly is, therefore, related with the possibility of 
having particles moving from one scale to the other as 
temperature is changed what becomes quite difficult if
the depth of the shoulder well becomes too deep.

\subsection*{Diffusion anomaly}

Now we are going to test the effect the shoulder depth has in the location of the diffusion anomaly in 
the pressure temperature phase diagram. The diffusion coefficient is obtained from the expression:
\begin{equation}
  D = \lim_{t \rightarrow \infty} \frac {\langle \left[ \vec{r}_j (t_0
      + t) - \vec{r}_j(t_0) \right]^2 \rangle_{t_0}} {6t}
  \label{eq:diffusion}
\end{equation}
where $\vec{r}_j(t)$ are the coordinates of particle $j$ at time $t$,
and $\langle \cdots \rangle_{t_0}$ denotes an average over all particles and over all $t_0$.

\begin{figure}[h]
  \begin{centering}
    \begin{tabular}{cc}
      \includegraphics[width=6cm]{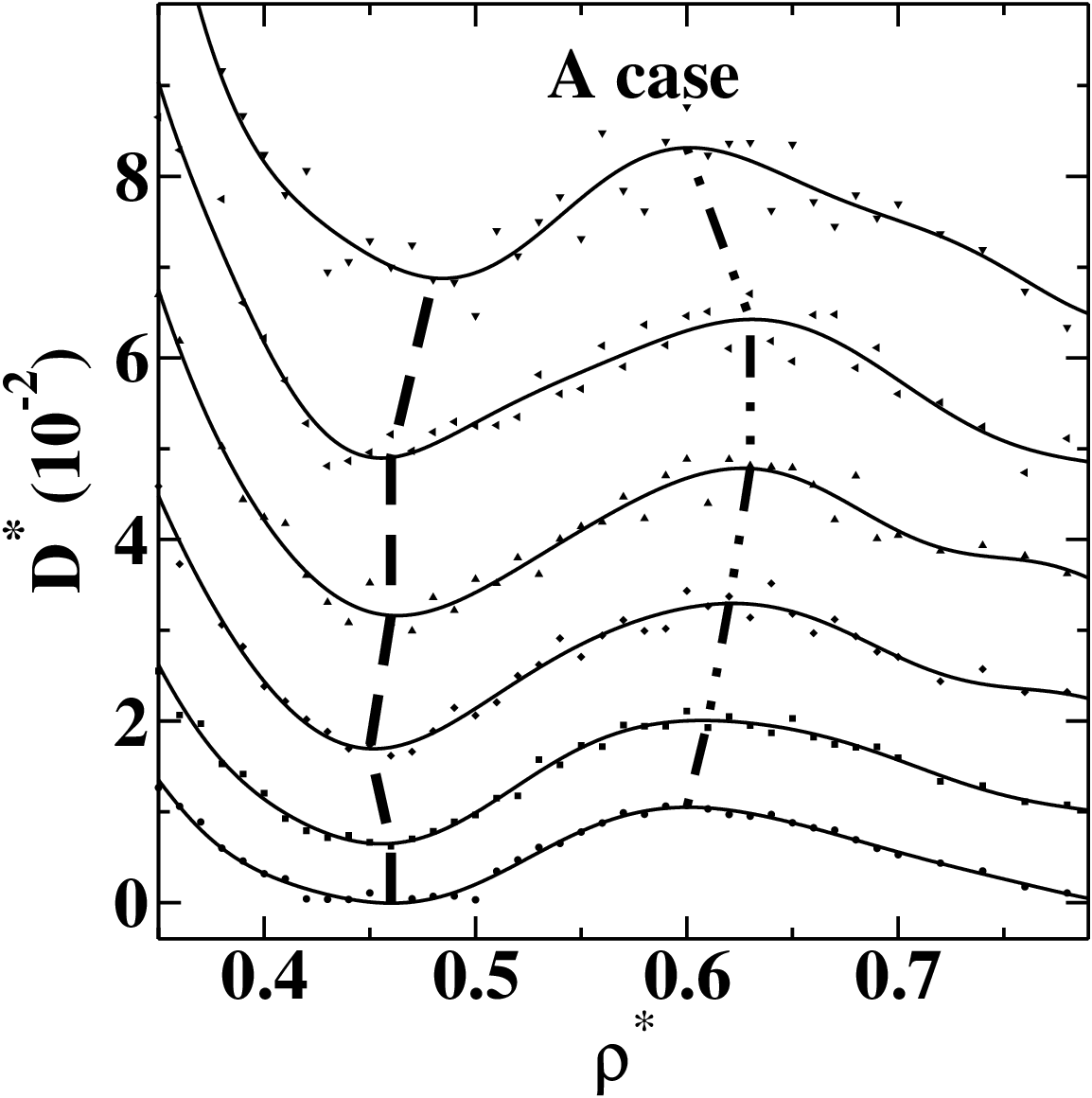} & \ 
      \includegraphics[width=5.75cm]{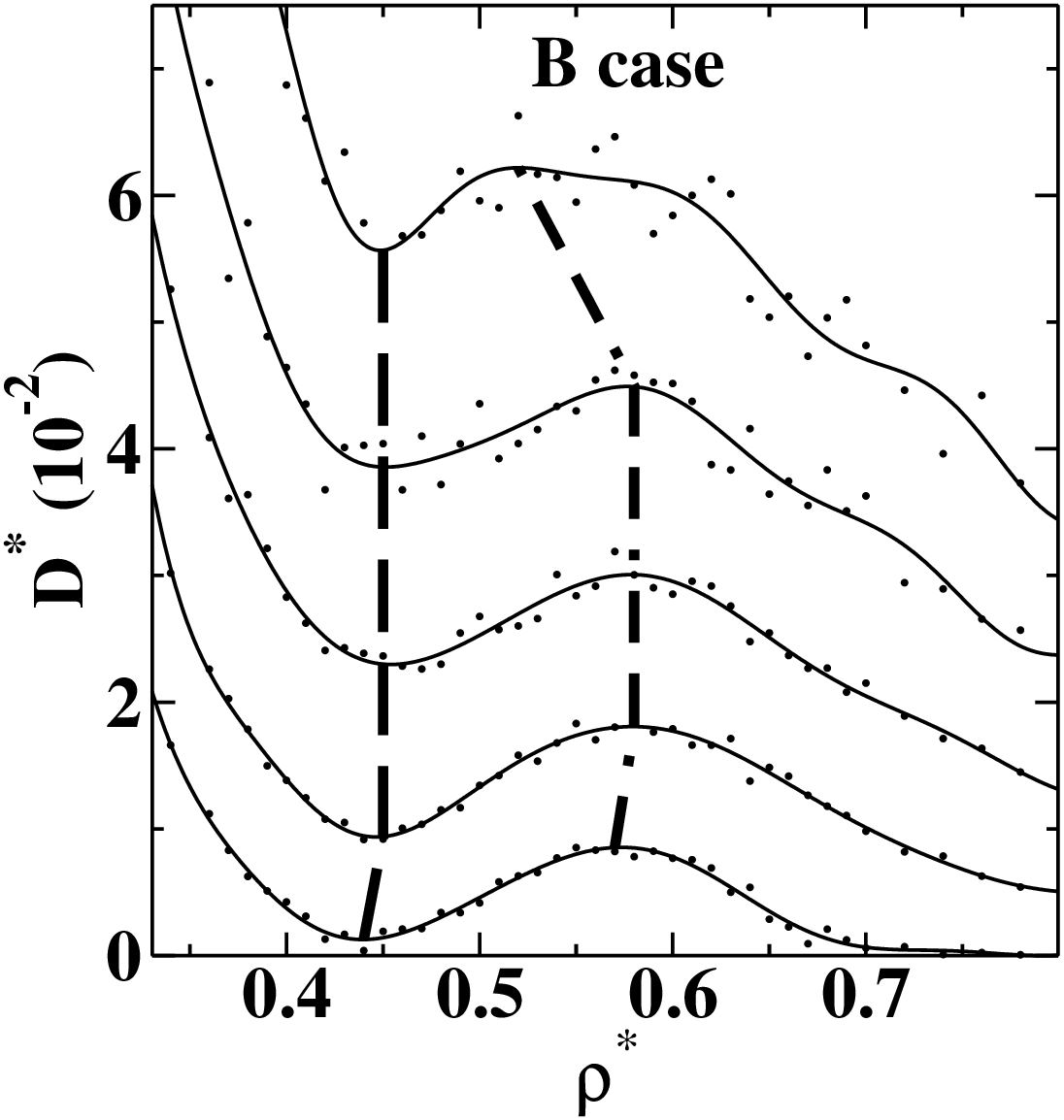} \tabularnewline
      \includegraphics[width=6cm]{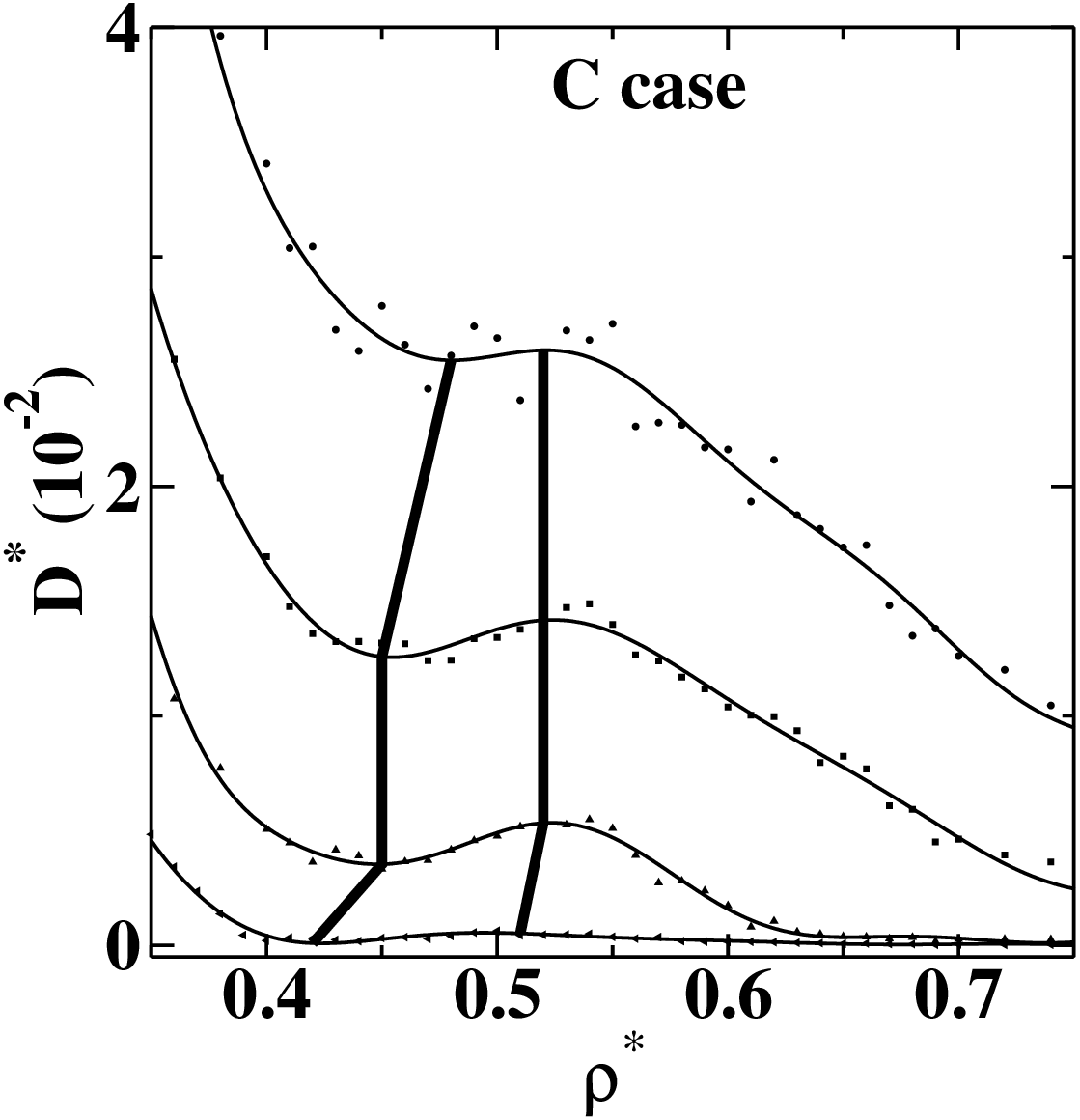} &
      \includegraphics[width=6.3cm]{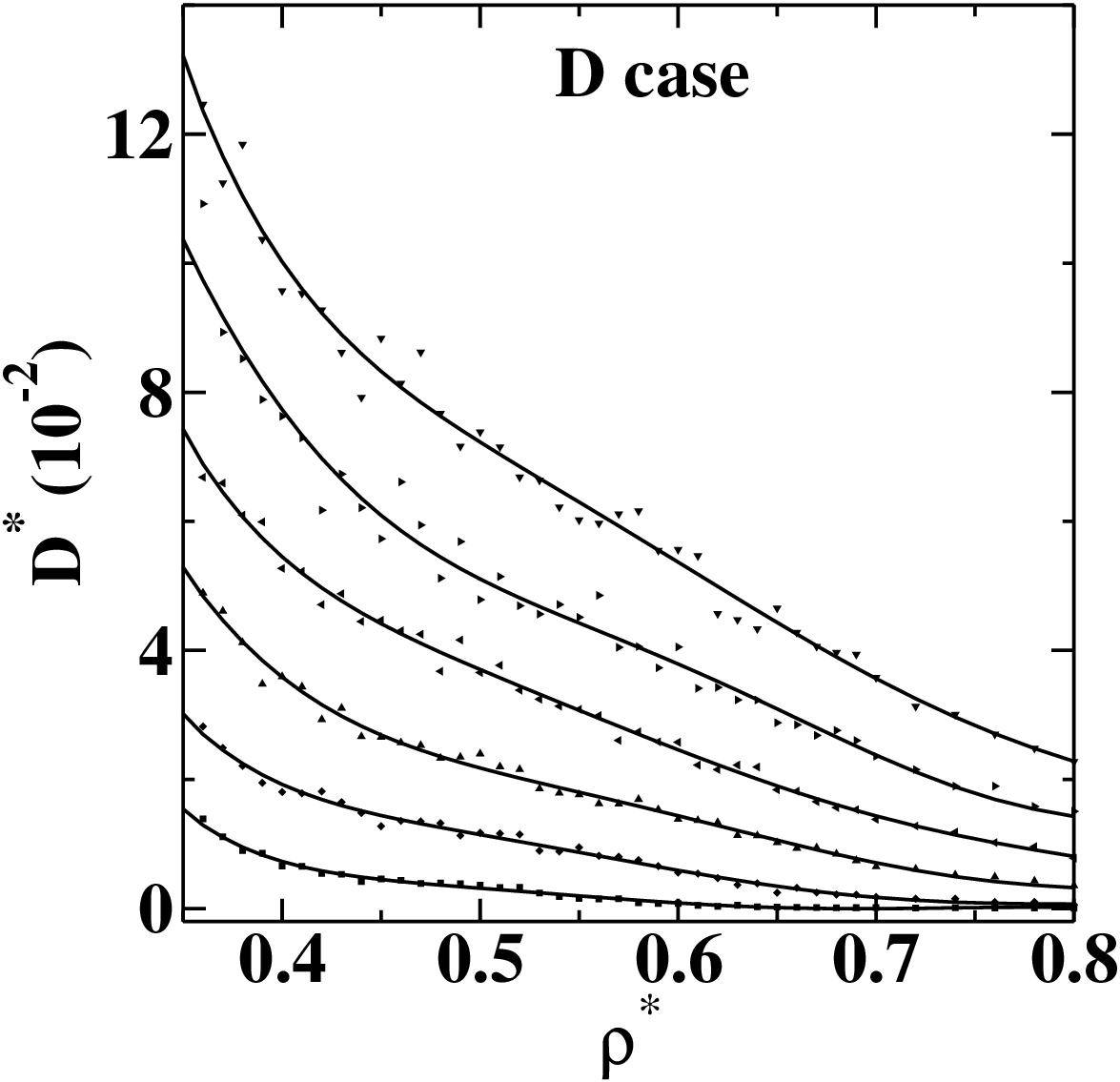} \tabularnewline
    \end{tabular}
    \par
  \end{centering}
  \caption{Diffusion coefficient as a function of density. The dots are the simulational data and the solid lines
are polynomial fits. The dashed lines connect the densities of  minima  and maxima diffusivity that
limit the diffusion anomalous region.}
  \label{fig:Drho}
\end{figure}

Fig.~\ref{fig:Drho} shows the behavior of the dimensionless translational 
diffusion coefficient, $D^*$, as function of the dimensionless 
density, $\rho^*$, at constant temperature  for the four cases. The
solid lines are a polynomial fits to the data obtained by simulation 
(the dots in the Fig.~\ref{fig:Drho}). For normal liquids, the 
diffusion at constant temperature increases with the decrease of 
the density. For the potentials $A$, $B$ and $C$ the diffusion 
has a region  in the 
pressure-temperature phase diagram where the diffusion increases 
with density what represents a diffusion anomalous 
region. In the Fig.~\ref{fig:Drho} one  dashed line joints the points 
of the density (or pressure) of minimum diffusion for different 
temperatures and another dashed line links the points of 
density (or pressure) of maximum diffusion for different temperatures.

Similarly to what happens with the location of the TMD,  as the 
shoulder becomes deeper, the region in the pressure-temperature phase 
diagram delimited by the extrema of the diffusion goes to lower 
pressures, shrinks and disappears for the case $D$, the potential 
with the deepest shoulder.

Fig.~\ref{fig:PT} shows the location at the pressure-temperature 
phase diagram of the pressure
of  maximum  and minimum diffusion as 
double dot dashed lines (the dashed lines in the 
Fig.~\ref{fig:Drho}). In the Fig.~\ref{fig:PT} we show that in the 
pressure-temperature phase diagram the region where the 
dynamic anomaly occurs englobes the region where the 
thermodynamic anomaly is present. This hierarchy between the 
anomalies is observed in a number of 
models~\cite{Ol08a,Er01,Ne01} and in the water.~\cite{An76}

\begin{figure}[h]
  \begin{centering}
    \begin{tabular}{cc} 
     \includegraphics[width=7.5cm]{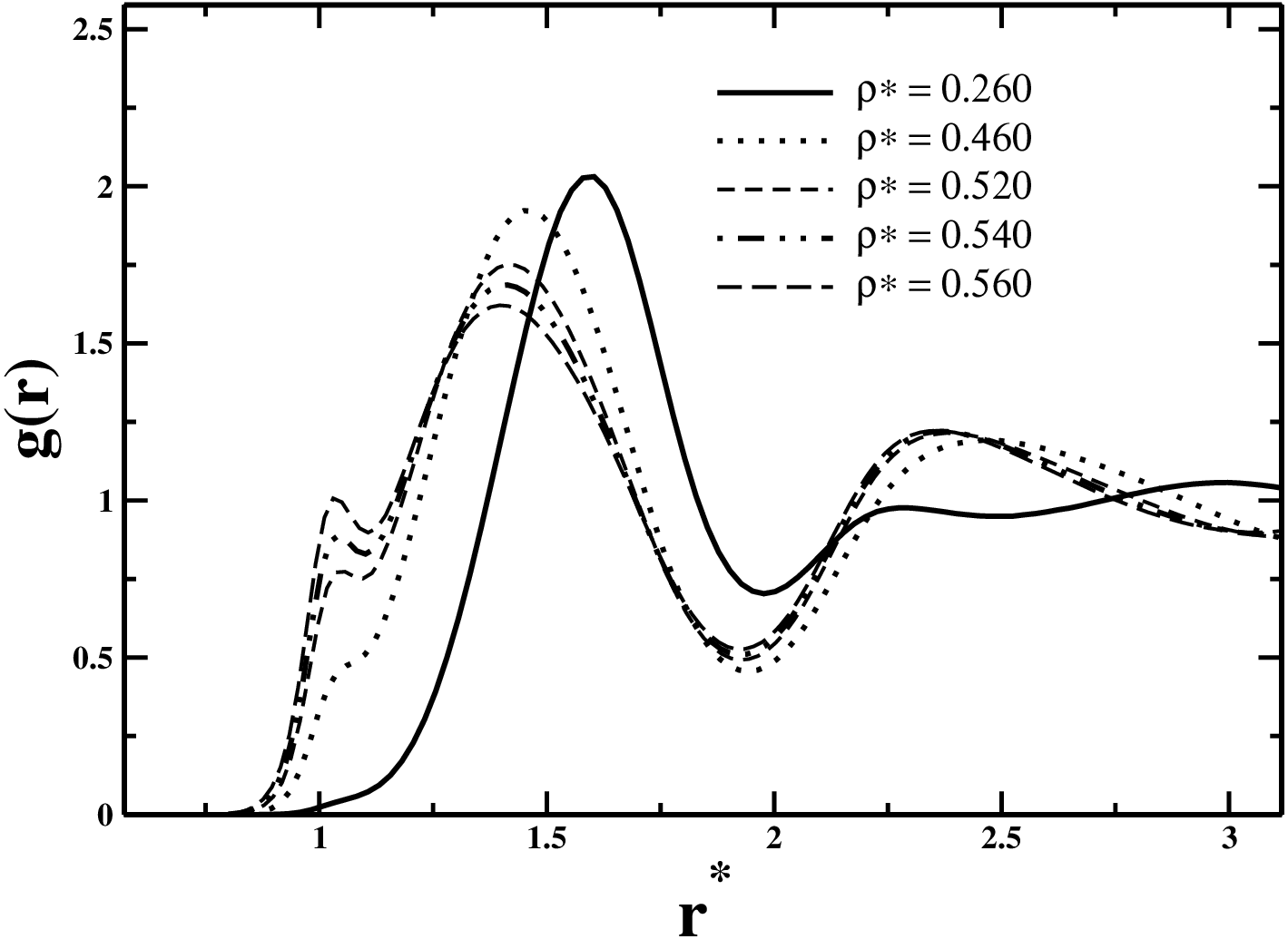} & \ 
     \includegraphics[width=7.7cm]{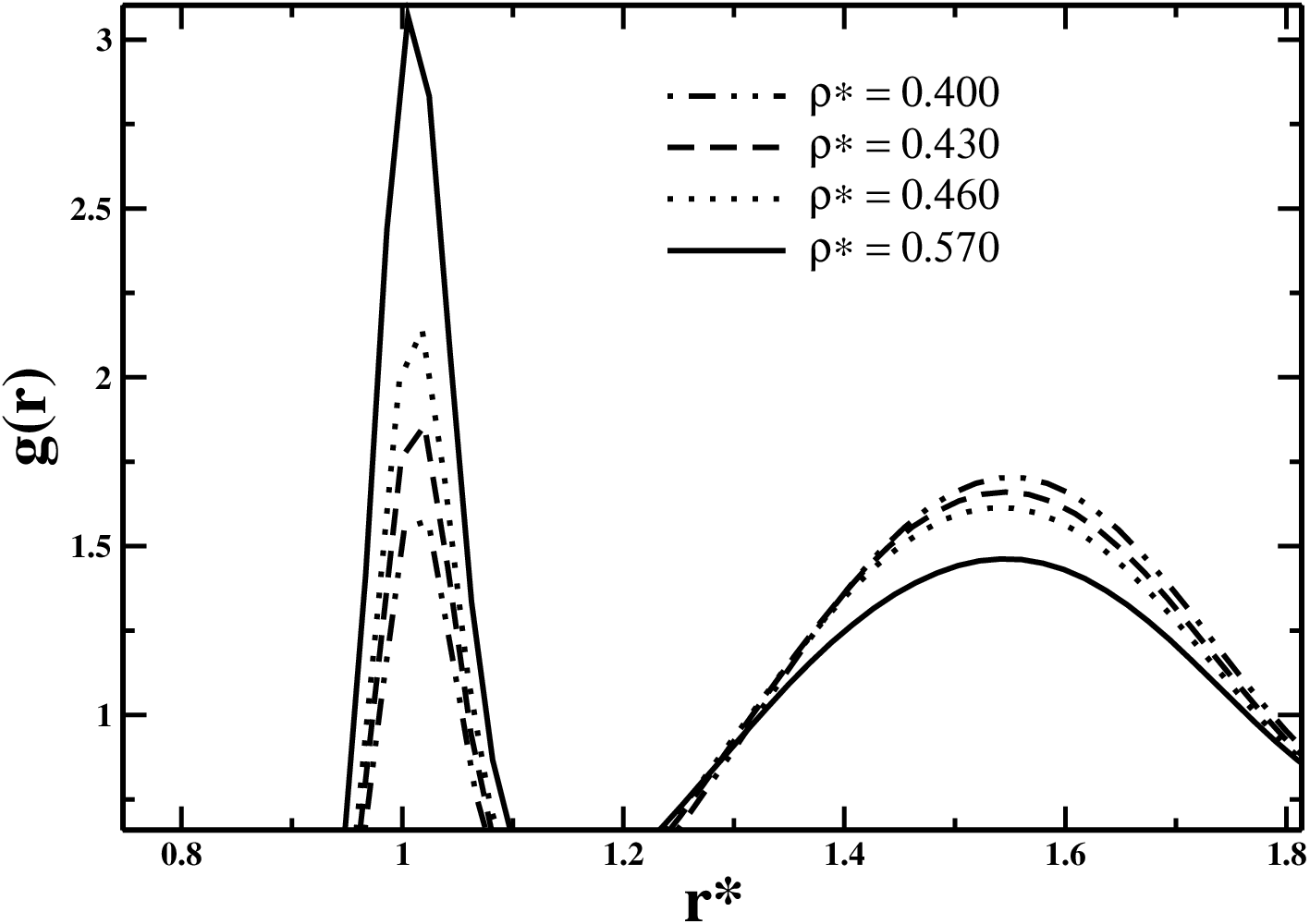} \tabularnewline\
    \end{tabular}
    \par
  \end{centering}
  \caption{Radial distribution for cases A and D as 
a function of $r^*$ for various densities. In the case A
the temperature is fixed $T^*=0.90$ while in the case D the
temperature is $T^*=1.10$. }
  \label{fig:gr2} 
\end{figure}

The link between the depth of the shoulder and the presence or 
not of the region of diffusion extrema goes as follows. 
The presence of the diffusion anomaly is related to having the
quantity $\Sigma_2 > 0.42$ \cite{Er06,Ol08a} where 
\begin{eqnarray}
\label{eq:sigma}
\Sigma_{2} & = & 
\left(\frac{\partial s_{2}}{\partial\ln\rho}\right)_{T}\\ \nonumber
 & = & s_{2}-2\pi\rho^{2}\int\ln g(r)\frac{\partial g(r)}{\partial\rho}r^{2}dr
\label{sigma}
\end{eqnarray}
where
\begin{equation}
s_{2}=-2\pi\rho\int\left[g(r)\ln g(r)-g(r)+1\right]r^2 dr
 \;,
\label{s2}
\end{equation}
is the excess entropy. Fig.~\ref{fig:gr2} illustrates the behavior
of the radial distribution function for fixed temperature as 
the density varies. For the case A the $\ln g(r)$ is negative and $dg(r)/d\rho$ is
positive for the first scale, while for the second scale the $\ln g(r)$ is positive and
the $dg(r)/d\rho$ is negative. As a result the second parcel in
Eq.~(\ref{sigma}) is positive a requirement for having $\Sigma_2 > 0.42$ since
$s_2$ is negative \cite{Ol08a}. For case D, also shown in
Fig.~\ref{fig:gr2}, the $\ln g(r)$ is positive and huge and  $dg(r)/d\rho$ is
positive what leads to a second parcel in Eq.~(\ref{sigma}) that
is negative what do not fulfill the requirement $\Sigma_2 > 0.42$.
If the shoulder well is too deep the particle is unable to
go from one scale to the other and the density anomalous behavior
does not happen.

\subsection*{Structural anomaly}

Finally we are going to test the effect the shoulder depth has in the location in the pressure-temperature phase diagram of the structural anomalous region.

\begin{figure}[h]
  \begin{centering}
    \begin{tabular}{cc}
      \includegraphics[width=6cm]{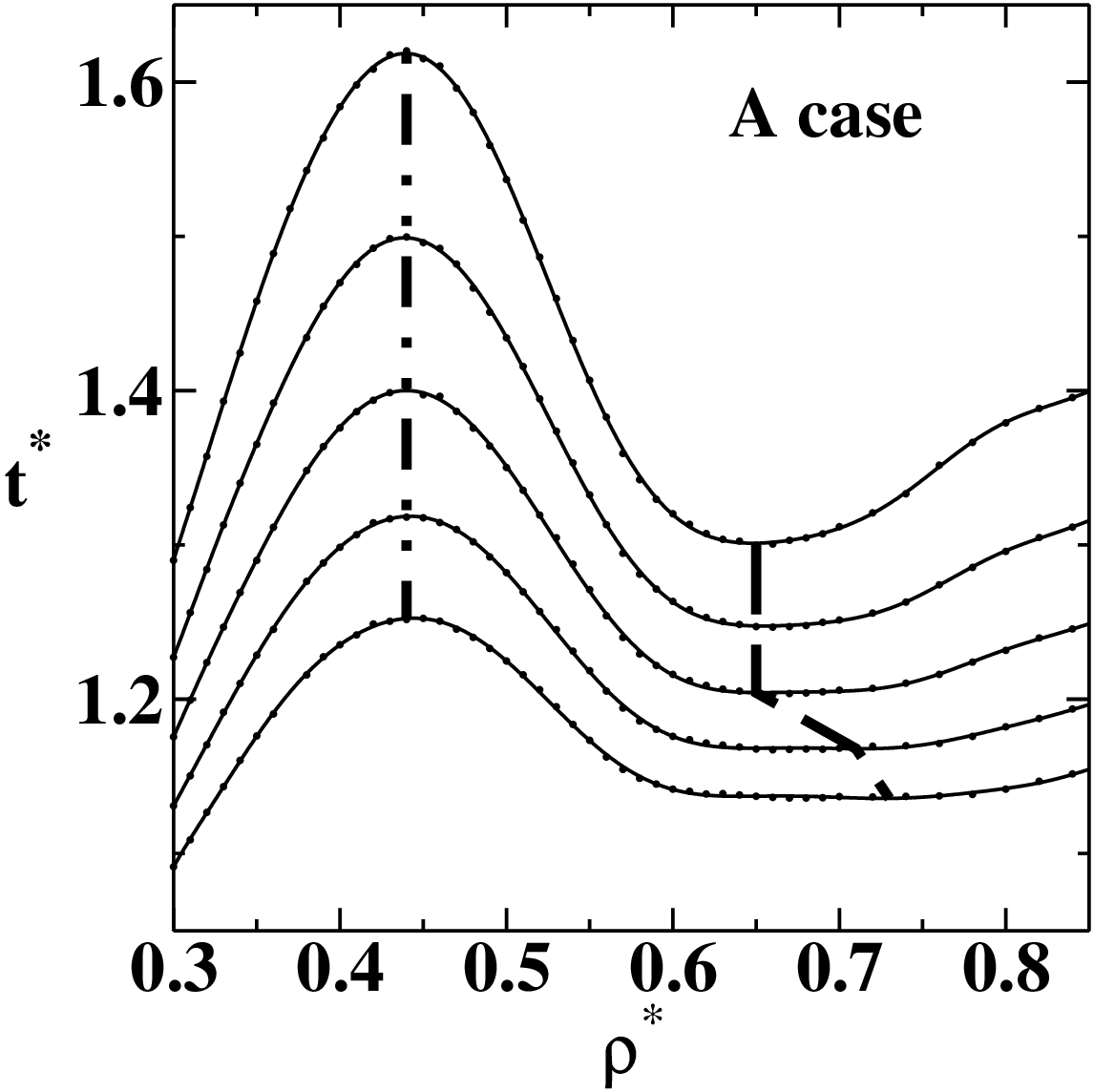} & \ 
      \includegraphics[width=6cm]{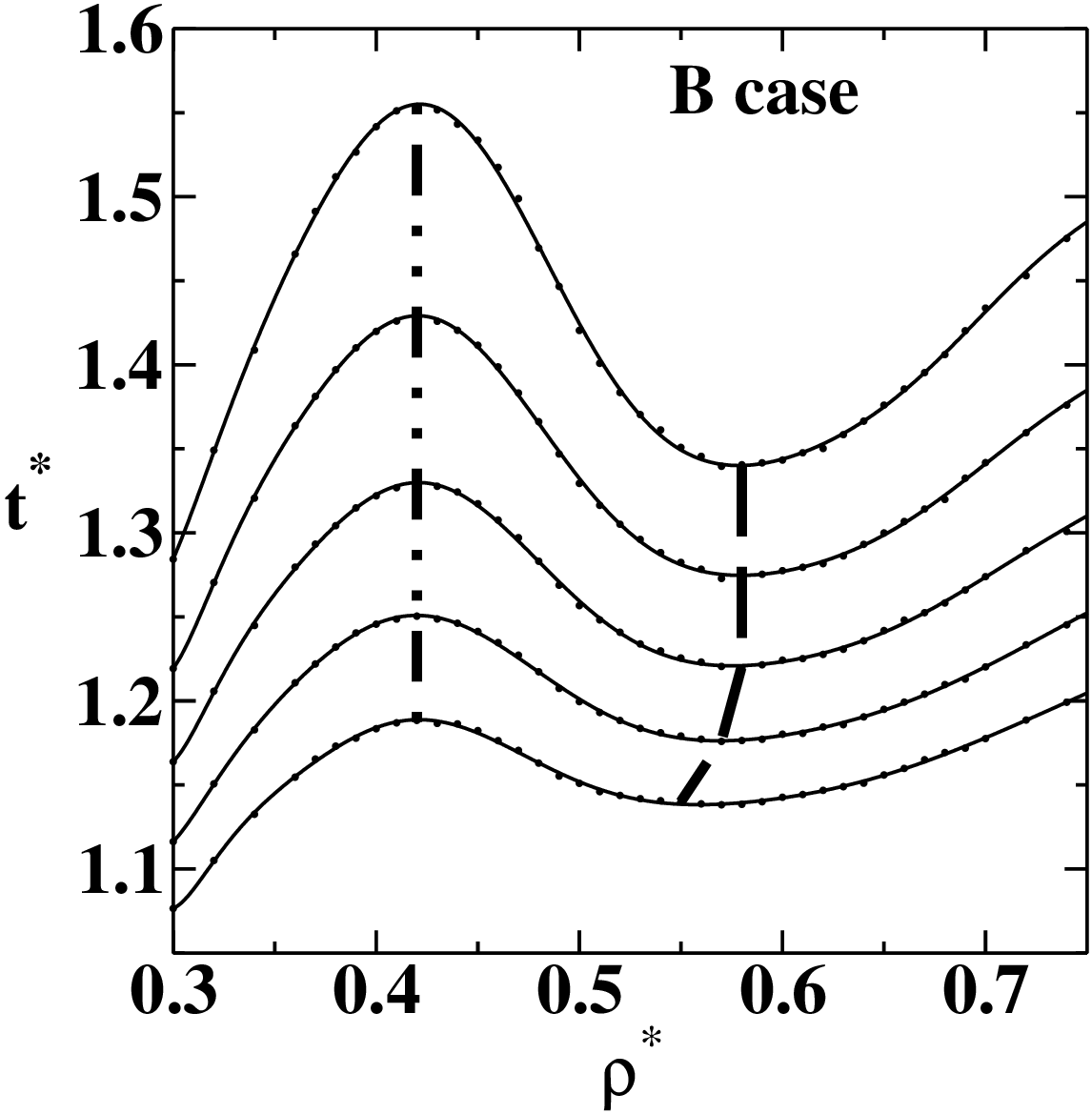} \tabularnewline
      \includegraphics[width=6cm]{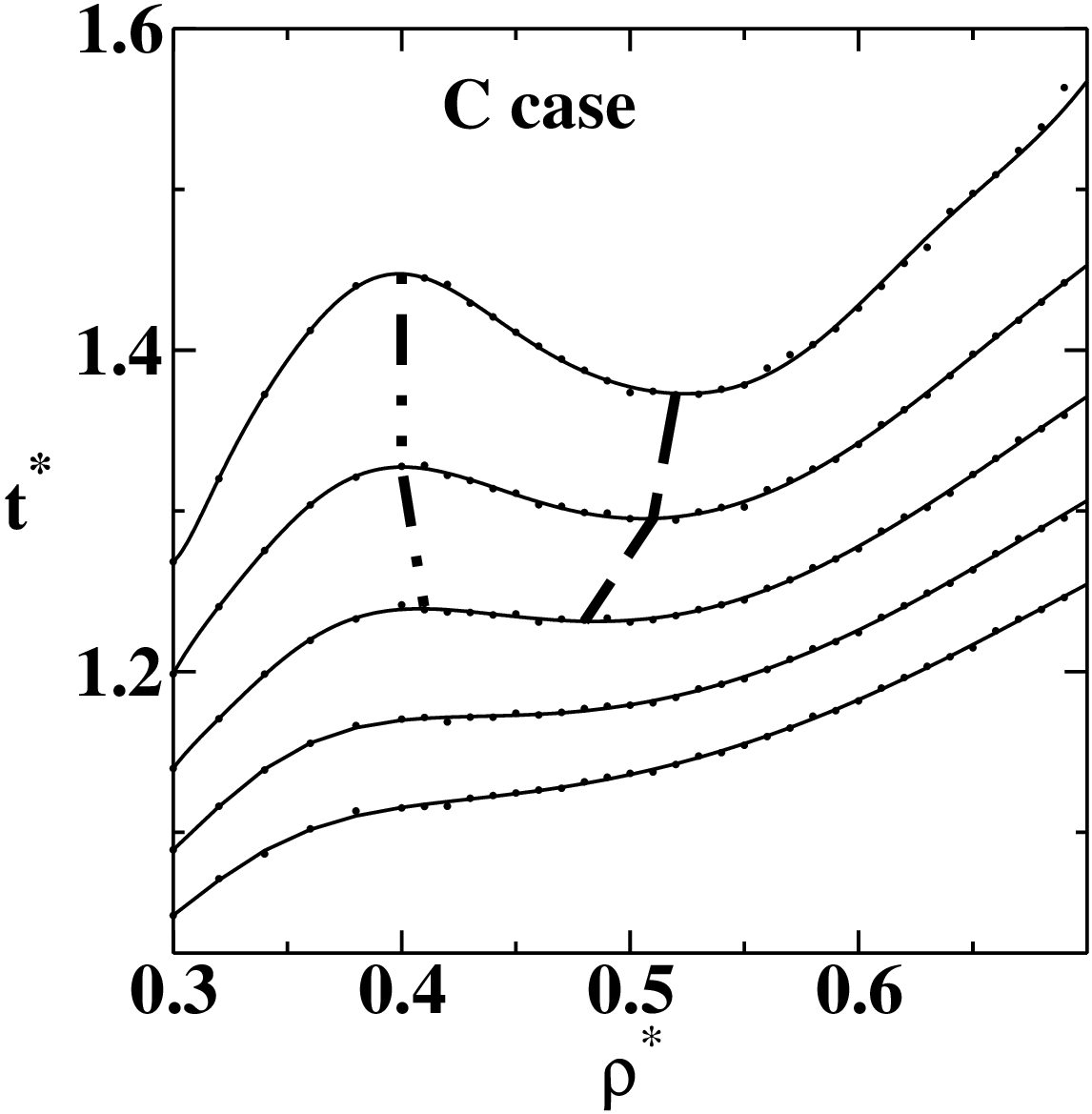} & \ 
      \includegraphics[width=6cm]{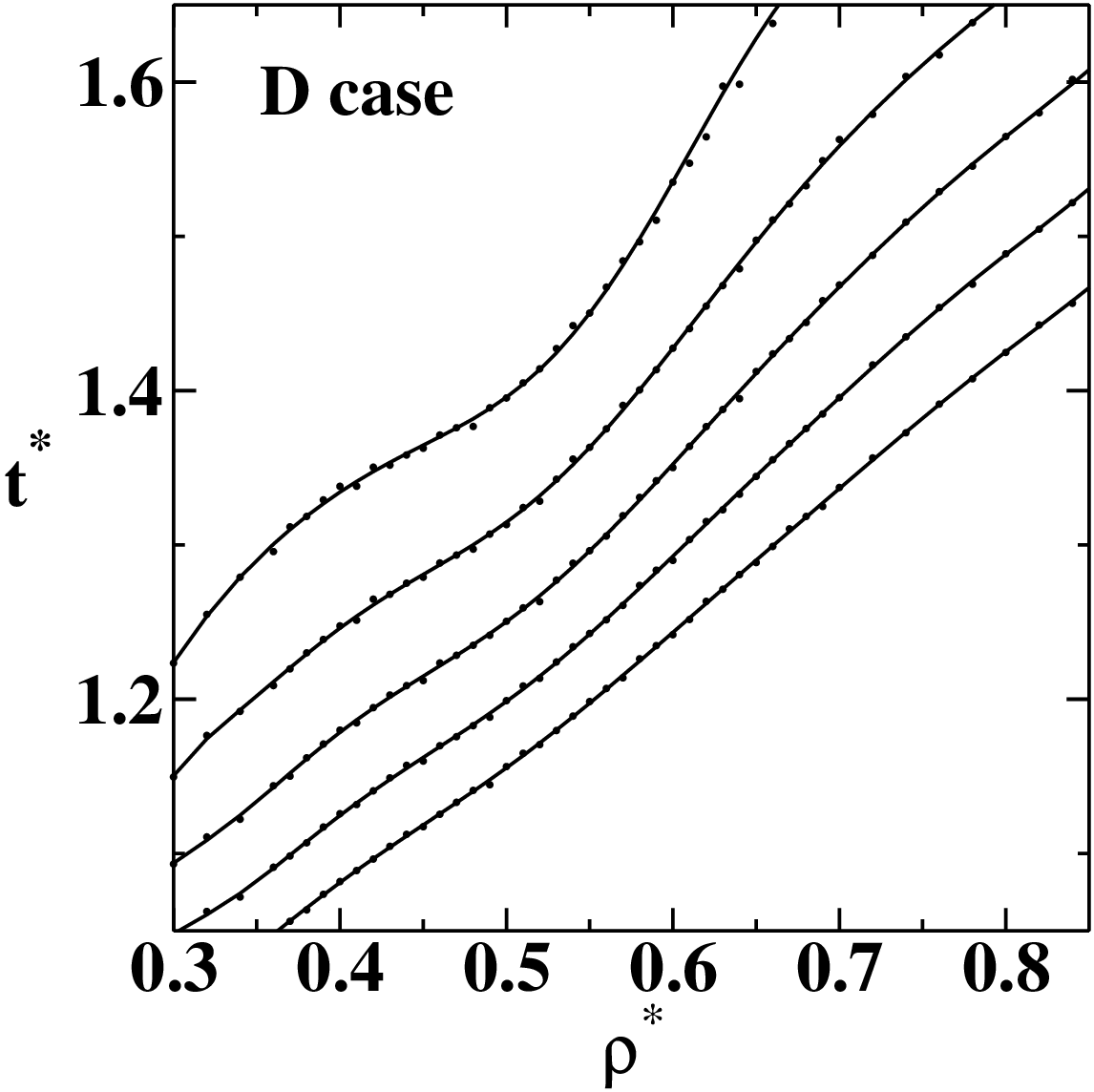} \tabularnewline
    \end{tabular}
    \par
  \end{centering}
  \caption{The translational order parameter as a function of density for fixed temperatures: $T^* = 1.10,\ 1.00,\ 0.90,\ 0.80,\ 0.70$ and $0.60$ (from top to bottom). The dot-dashed lines locate the density of maxima e minima $t^*$.}
  \label{fig:trho-1} 
\end{figure}

The translational order parameter is defined as~\cite{Sh02,Er01,Er03}

\begin{equation}
  t = \int_0^{\xi_c} \left| g(\xi) - 1 \right| d\xi
  \label{eq_trans}
\end{equation}
where $\xi = r \rho^{\frac{1}{3}}$ is the distance $r$ in units of the
mean interparticle separation $\rho^{-\frac{1}{3}}$, $\xi_{c}$ is the
cutoff distance set to half of the simulation box times~\cite{Ol06b}
$\rho^{-\frac{1}{3}}$, $g(\xi)$ is the radial distribution function
proportional to the probability of finding a particle at a distance
$\xi$ from a referent particle. The translational
order parameter measure how structured is
the system. For an ideal gas $g=1$ and $t=0$, and
the case of crystal phase $g\neq1$ over long distances and $t$ is large.
Therefore for normal fluids $t$ increases with the increase of 
the density.

Fig.~\ref{fig:trho-1} shows  the translational order parameter as a function of 
density for fixed temperature. The dots represent the simulation data and the solid line the  polynomial fit to the data. For the potentials $A$, $B$ and $C$ there are a region of densities in which the translational parameter decreases as the  density increases. A dotted-dashed line illustrates the region of local maximum of $t^*$ and 
minimum of $t^*$ limiting the anomalous region. For the potential $D$,  $t^*$ increases with the density. No anomalous behavior is observed.

Fig.~\ref{fig:PT} shows the structural anomaly for cases $A$, $B$ and $C$, as dashed-pointed (purple) lines. It is observed that the region of structural anomaly embraces both dynamic and thermodynamic anomalies. Similarly to other anomalies the effect of increase the depth of the repulsive shoulder is to narrow the anomalies asymmetrical. The branch of anomaly in pressures near to liquid-liquid critical point is most feeling to the effect of the shoulder compared with the branch obtained in low pressures. However, the hierarchy of the anomalies is maintained, the
change in the repulsive shoulder does not affect it.

\section{\label{sec4} Conclusions}

In this paper we studied a family of potentials characterized by  two length scales: a repulsive shoulder and 
an attractive well. We analyzed the effect  in the location in the pressure-temperature phase diagram  of the 
density, diffusion and structural anomalies of  making this repulsive shoulder a deep  well. We found that the anomalies shrink and disappear as the well becomes deeper. This indicates that an important mechanism for the anomalies is the possibility of particles in the furthest length scale to move to the
closest  length scale. As the shoulder well becomes deeper particles becomes localized in the closest scale and the mobility between the two scales decreases.

We find that in the cases of potentials $A$, $B$ and $C$ the thermodynamic, dynamic an structural anomalies are present and that  the region of structural anomaly embraces the dynamic and thermodynamic anomaly in pressure-temperature phase diagram. This implies that the hierarchy of the anomalies is preserved independent of the depth of the repulsive shoulder, however when the shoulder becomes deeper, the upper pressure lines of anomaly
converge to a similar value in the pressure-temperature phase diagram.

What is the connection between the studies potentials and the real system? Effective potentials for water has been derived based in the oxygen-oxygen radial distribution function for the ST4~\cite{He93} and TIP5P~\cite{Ya08} models for water. In both cases the effective potential was obtained from the g$(r^*)$ using the Ornstein-Zernike equation and integral equation approximations. The potential resulting  are  the case $D$ in the Fig.~\ref{fig:potential} in 
the case of ST4 and  for the TIP5P model a potential that exhibits a deep shoulder similar to the case $D$.
Consequently the approximation washes out the anomalies present in both ST4 and TIP5P. In the case of the TIP5P it was shown that if instead of deep shoulder a smooth shoulder like the one present in the ramp potential would
be used, the anomalies not only would be present but would be located in the same region of pressure and temperature of the TIP5P potential.

In resume, similarly to other previous studies~\cite{Ol06a,Wi02,Ca03,La98}, a directional 
interaction potential is not a fundamental ingredient to have a water-like anomalies. Two scales isotropic
potential also reproduce this anomalies if the shoulder closest scale would not be too deep.

\section*{ACKNOWLEDGMENTS}

We thank for financial support the Brazilian science agencies CNPq and Capes. This work is partially supported by CNPq, INCT-FCx.

\vspace{1cm}

\bibliography{Biblioteca}

\begin{thebibliography}{59}
\expandafter\ifx\csname natexlab\endcsname\relax\def\natexlab#1{#1}\fi
\expandafter\ifx\csname bibnamefont\endcsname\relax
  \def\bibnamefont#1{#1}\fi
\expandafter\ifx\csname bibfnamefont\endcsname\relax
  \def\bibfnamefont#1{#1}\fi
\expandafter\ifx\csname citenamefont\endcsname\relax
  \def\citenamefont#1{#1}\fi
\expandafter\ifx\csname url\endcsname\relax
  \def\url#1{\texttt{#1}}\fi
\expandafter\ifx\csname urlprefix\endcsname\relax\def\urlprefix{URL }\fi
\providecommand{\bibinfo}[2]{#2}
\providecommand{\eprint}[2][]{\url{#2}}

\bibitem[{\citenamefont{Waler}(1964)}]{Wa64}
\bibinfo{author}{\bibfnamefont{R.}~\bibnamefont{Waler}},
  \emph{\bibinfo{title}{Essays of natural experiments}}
  (\bibinfo{publisher}{Johnson Reprint}, \bibinfo{address}{New York},
  \bibinfo{year}{1964}).

\bibitem[{\citenamefont{Angell et~al.}(1976)\citenamefont{Angell, Finch, and
  Bach}}]{An76}
\bibinfo{author}{\bibfnamefont{C.~A.} \bibnamefont{Angell}},
  \bibinfo{author}{\bibfnamefont{E.~D.} \bibnamefont{Finch}}, \bibnamefont{and}
  \bibinfo{author}{\bibfnamefont{P.}~\bibnamefont{Bach}}, \bibinfo{journal}{J.
  Chem. Phys.} \textbf{\bibinfo{volume}{65}}, \bibinfo{pages}{3063}
  (\bibinfo{year}{1976}).

\bibitem[{\citenamefont{Prielmeier et~al.}(1987)\citenamefont{Prielmeier, Lang,
  Speedy, and L\"udemann}}]{Pr87}
\bibinfo{author}{\bibfnamefont{F.~X.} \bibnamefont{Prielmeier}},
  \bibinfo{author}{\bibfnamefont{E.~W.} \bibnamefont{Lang}},
  \bibinfo{author}{\bibfnamefont{R.~J.} \bibnamefont{Speedy}},
  \bibnamefont{and} \bibinfo{author}{\bibfnamefont{H.-D.}
  \bibnamefont{L\"udemann}}, \bibinfo{journal}{Phys. Rev. Lett.}
  \textbf{\bibinfo{volume}{59}}, \bibinfo{pages}{1128} (\bibinfo{year}{1987}).

\bibitem[{\citenamefont{Prielmeier et~al.}(1998)\citenamefont{Prielmeier, Lang,
  Speedy, and L\"udemann}}]{Pr88}
\bibinfo{author}{\bibfnamefont{F.~X.} \bibnamefont{Prielmeier}},
  \bibinfo{author}{\bibfnamefont{E.~W.} \bibnamefont{Lang}},
  \bibinfo{author}{\bibfnamefont{R.~J.} \bibnamefont{Speedy}},
  \bibnamefont{and} \bibinfo{author}{\bibfnamefont{H.-D.}
  \bibnamefont{L\"udemann}}, \bibinfo{journal}{Ber. Bunsenges. Phys. Chem.}
  \textbf{\bibinfo{volume}{92}}, \bibinfo{pages}{1111} (\bibinfo{year}{1998}).

\bibitem[{\citenamefont{Haar et~al.}(1984)\citenamefont{Haar, Gallangher, and
  Kell}}]{Ha84}
\bibinfo{author}{\bibfnamefont{L.}~\bibnamefont{Haar}},
  \bibinfo{author}{\bibfnamefont{J.~S.} \bibnamefont{Gallangher}},
  \bibnamefont{and} \bibinfo{author}{\bibfnamefont{G.}~\bibnamefont{Kell}},
  \emph{\bibinfo{title}{NBS/NRC Steam Tables. Thermodyanic and Transport
  Properties and Computer Programs for Vapor and Liquid States of Water in SI
  Units.}} (\bibinfo{publisher}{Hemisphere Publishing Co.},
  \bibinfo{address}{Washington D. C.}, \bibinfo{year}{1984}),
  \bibinfo{edition}{1st} ed.

\bibitem[{\citenamefont{Thurn and Ruska}(1976)}]{Th76}
\bibinfo{author}{\bibfnamefont{H.}~\bibnamefont{Thurn}} \bibnamefont{and}
  \bibinfo{author}{\bibfnamefont{J.}~\bibnamefont{Ruska}}, \bibinfo{journal}{J.
  Non-Cryst. Solids} \textbf{\bibinfo{volume}{22}}, \bibinfo{pages}{331}
  (\bibinfo{year}{1976}).

\bibitem[{Los(2007)}]{LosAlamos}
\emph{\bibinfo{title}{Periodic table of the elements}},
  \bibinfo{howpublished}{\url{http://periodic.lanl.gov/default.htm}}
  (\bibinfo{year}{2007}).

\bibitem[{\citenamefont{Sauer and Borst}(1967)}]{Sa67}
\bibinfo{author}{\bibfnamefont{G.~E.} \bibnamefont{Sauer}} \bibnamefont{and}
  \bibinfo{author}{\bibfnamefont{L.~B.} \bibnamefont{Borst}},
  \bibinfo{journal}{Science} \textbf{\bibinfo{volume}{158}},
  \bibinfo{pages}{1567} (\bibinfo{year}{1967}).

\bibitem[{\citenamefont{Kennedy and Wheeler}(1983)}]{Ke83}
\bibinfo{author}{\bibfnamefont{S.~J.} \bibnamefont{Kennedy}} \bibnamefont{and}
  \bibinfo{author}{\bibfnamefont{J.~C.} \bibnamefont{Wheeler}},
  \bibinfo{journal}{J. Chem. Phys.} \textbf{\bibinfo{volume}{78}},
  \bibinfo{pages}{1523} (\bibinfo{year}{1983}).

\bibitem[{\citenamefont{Tsuchiya}(1991)}]{Ts91}
\bibinfo{author}{\bibfnamefont{T.}~\bibnamefont{Tsuchiya}},
  \bibinfo{journal}{J. Phys. Soc. Jpn.} \textbf{\bibinfo{volume}{60}},
  \bibinfo{pages}{227} (\bibinfo{year}{1991}).

\bibitem[{\citenamefont{Angell et~al.}(2000)\citenamefont{Angell, Bressel,
  Hemmatti, Sare, and Tucker}}]{An00}
\bibinfo{author}{\bibfnamefont{C.~A.} \bibnamefont{Angell}},
  \bibinfo{author}{\bibfnamefont{R.~D.} \bibnamefont{Bressel}},
  \bibinfo{author}{\bibfnamefont{M.}~\bibnamefont{Hemmatti}},
  \bibinfo{author}{\bibfnamefont{E.~J.} \bibnamefont{Sare}}, \bibnamefont{and}
  \bibinfo{author}{\bibfnamefont{J.~C.} \bibnamefont{Tucker}},
  \bibinfo{journal}{Phys. Chem. Chem. Phys.} \textbf{\bibinfo{volume}{2}},
  \bibinfo{pages}{1559} (\bibinfo{year}{2000}).

\bibitem[{\citenamefont{Sharma et~al.}(2006)\citenamefont{Sharma, Chakraborty,
  and Chakravarty}}]{Ru06b}
\bibinfo{author}{\bibfnamefont{R.}~\bibnamefont{Sharma}},
  \bibinfo{author}{\bibfnamefont{S.~N.} \bibnamefont{Chakraborty}},
  \bibnamefont{and}
  \bibinfo{author}{\bibfnamefont{C.}~\bibnamefont{Chakravarty}},
  \bibinfo{journal}{J. Chem. Phys.} \textbf{\bibinfo{volume}{125}},
  \bibinfo{pages}{204501} (\bibinfo{year}{2006}).

\bibitem[{\citenamefont{Shell et~al.}(2002)\citenamefont{Shell, Debenedetti,
  and Panagiotopoulos}}]{Sh02}
\bibinfo{author}{\bibfnamefont{M.~S.} \bibnamefont{Shell}},
  \bibinfo{author}{\bibfnamefont{P.~G.} \bibnamefont{Debenedetti}},
  \bibnamefont{and} \bibinfo{author}{\bibfnamefont{A.~Z.}
  \bibnamefont{Panagiotopoulos}}, \bibinfo{journal}{Phys. Rev. E}
  \textbf{\bibinfo{volume}{66}}, \bibinfo{pages}{011202}
  (\bibinfo{year}{2002}).

\bibitem[{\citenamefont{Poole et~al.}(1997)\citenamefont{Poole, Hemmati, and
  Angell}}]{Po97}
\bibinfo{author}{\bibfnamefont{P.~H.} \bibnamefont{Poole}},
  \bibinfo{author}{\bibfnamefont{M.}~\bibnamefont{Hemmati}}, \bibnamefont{and}
  \bibinfo{author}{\bibfnamefont{C.~A.} \bibnamefont{Angell}},
  \bibinfo{journal}{Phys. Rev. Lett.} \textbf{\bibinfo{volume}{79}},
  \bibinfo{pages}{2281} (\bibinfo{year}{1997}).

\bibitem[{\citenamefont{Sastry and Angell}(2003)}]{Sa03}
\bibinfo{author}{\bibfnamefont{S.}~\bibnamefont{Sastry}} \bibnamefont{and}
  \bibinfo{author}{\bibfnamefont{C.~A.} \bibnamefont{Angell}},
  \bibinfo{journal}{Nature Mater.} \textbf{\bibinfo{volume}{2}},
  \bibinfo{pages}{739} (\bibinfo{year}{2003}).

\bibitem[{\citenamefont{Berendsen et~al.}(1987)\citenamefont{Berendsen,
  Grigera, and Straatsma}}]{spce}
\bibinfo{author}{\bibfnamefont{H.~J.~C.} \bibnamefont{Berendsen}},
  \bibinfo{author}{\bibfnamefont{J.~R.} \bibnamefont{Grigera}},
  \bibnamefont{and} \bibinfo{author}{\bibfnamefont{T.~P.}
  \bibnamefont{Straatsma}}, \bibinfo{journal}{J. Phys. Chem.}
  \textbf{\bibinfo{volume}{91}}, \bibinfo{pages}{6269} (\bibinfo{year}{1987}).

\bibitem[{\citenamefont{Netz et~al.}(2001)\citenamefont{Netz, Starr, Stanley,
  and Barbosa}}]{Ne01}
\bibinfo{author}{\bibfnamefont{P.~A.} \bibnamefont{Netz}},
  \bibinfo{author}{\bibfnamefont{F.~W.} \bibnamefont{Starr}},
  \bibinfo{author}{\bibfnamefont{H.~E.} \bibnamefont{Stanley}},
  \bibnamefont{and} \bibinfo{author}{\bibfnamefont{M.~C.}
  \bibnamefont{Barbosa}}, \bibinfo{journal}{J. Chem. Phys.}
  \textbf{\bibinfo{volume}{115}}, \bibinfo{pages}{344} (\bibinfo{year}{2001}).

\bibitem[{\citenamefont{Errington and Debenedetti}(2001)}]{Er01}
\bibinfo{author}{\bibfnamefont{J.~R.} \bibnamefont{Errington}}
  \bibnamefont{and} \bibinfo{author}{\bibfnamefont{P.~G.}
  \bibnamefont{Debenedetti}}, \bibinfo{journal}{Nature (London)}
  \textbf{\bibinfo{volume}{409}}, \bibinfo{pages}{318} (\bibinfo{year}{2001}).

\bibitem[{\citenamefont{Mittal et~al.}(2006)\citenamefont{Mittal, Errington,
  and Truskett}}]{Mi06a}
\bibinfo{author}{\bibfnamefont{J.}~\bibnamefont{Mittal}},
  \bibinfo{author}{\bibfnamefont{J.~R.} \bibnamefont{Errington}},
  \bibnamefont{and} \bibinfo{author}{\bibfnamefont{T.~M.}
  \bibnamefont{Truskett}}, \bibinfo{journal}{J. Phys. Chem. B}
  \textbf{\bibinfo{volume}{110}}, \bibinfo{pages}{18147}
  (\bibinfo{year}{2006}).

\bibitem[{\citenamefont{Kumar et~al.}(2006)\citenamefont{Kumar, Franzese, and
  Stanley}}]{Ku06}
\bibinfo{author}{\bibfnamefont{P.}~\bibnamefont{Kumar}},
  \bibinfo{author}{\bibfnamefont{G.}~\bibnamefont{Franzese}}, \bibnamefont{and}
  \bibinfo{author}{\bibfnamefont{H.~E.} \bibnamefont{Stanley}},
  \bibinfo{journal}{Phys. Rev. E} \textbf{\bibinfo{volume}{73}},
  \bibinfo{pages}{041505} (\bibinfo{year}{2006}).

\bibitem[{\citenamefont{Mudi et~al.}(2005)\citenamefont{Mudi, Chakravarty, and
  Ramaswamy}}]{Mu05}
\bibinfo{author}{\bibfnamefont{A.}~\bibnamefont{Mudi}},
  \bibinfo{author}{\bibfnamefont{C.}~\bibnamefont{Chakravarty}},
  \bibnamefont{and}
  \bibinfo{author}{\bibfnamefont{R.}~\bibnamefont{Ramaswamy}},
  \bibinfo{journal}{J. Chem. Phys.} \textbf{\bibinfo{volume}{122}},
  \bibinfo{pages}{104507} (\bibinfo{year}{2005}).

\bibitem[{\citenamefont{Chen et~al.}(2006)\citenamefont{Chen, Mallamace, Mou,
  Broccio, Corsaro, Faraone, and Liu}}]{Ch06}
\bibinfo{author}{\bibfnamefont{S.~H.} \bibnamefont{Chen}},
  \bibinfo{author}{\bibfnamefont{F.}~\bibnamefont{Mallamace}},
  \bibinfo{author}{\bibfnamefont{C.~Y.} \bibnamefont{Mou}},
  \bibinfo{author}{\bibfnamefont{M.}~\bibnamefont{Broccio}},
  \bibinfo{author}{\bibfnamefont{C.}~\bibnamefont{Corsaro}},
  \bibinfo{author}{\bibfnamefont{A.}~\bibnamefont{Faraone}}, \bibnamefont{and}
  \bibinfo{author}{\bibfnamefont{L.}~\bibnamefont{Liu}},
  \bibinfo{journal}{Proceedings of the National Academy of Science of United
  States of America} \textbf{\bibinfo{volume}{103}}, \bibinfo{pages}{12974}
  (\bibinfo{year}{2006}).

\bibitem[{\citenamefont{Morishita}(2005)}]{Mo05}
\bibinfo{author}{\bibfnamefont{T.}~\bibnamefont{Morishita}},
  \bibinfo{journal}{Phys. Rev. E} \textbf{\bibinfo{volume}{72}},
  \bibinfo{pages}{021201} (\bibinfo{year}{2005}).

\bibitem[{\citenamefont{Poole et~al.}(1992)\citenamefont{Poole, Sciortino,
  Essmann, and Stanley}}]{Po92}
\bibinfo{author}{\bibfnamefont{P.~H.} \bibnamefont{Poole}},
  \bibinfo{author}{\bibfnamefont{F.}~\bibnamefont{Sciortino}},
  \bibinfo{author}{\bibfnamefont{U.}~\bibnamefont{Essmann}}, \bibnamefont{and}
  \bibinfo{author}{\bibfnamefont{H.~E.} \bibnamefont{Stanley}},
  \bibinfo{journal}{Nature (London)} \textbf{\bibinfo{volume}{360}},
  \bibinfo{pages}{324} (\bibinfo{year}{1992}).

\bibitem[{\citenamefont{Mishima and Stanley}(1998)}]{Mi98}
\bibinfo{author}{\bibfnamefont{O.}~\bibnamefont{Mishima}} \bibnamefont{and}
  \bibinfo{author}{\bibfnamefont{H.~E.} \bibnamefont{Stanley}},
  \bibinfo{journal}{Nature (London)} \textbf{\bibinfo{volume}{396}},
  \bibinfo{pages}{329} (\bibinfo{year}{1998}).

\bibitem[{\citenamefont{Speedy and Angell}(1976)}]{Sp76}
\bibinfo{author}{\bibfnamefont{R.~J.} \bibnamefont{Speedy}} \bibnamefont{and}
  \bibinfo{author}{\bibfnamefont{C.~A.} \bibnamefont{Angell}},
  \bibinfo{journal}{Journal of Chem. Phys.} \textbf{\bibinfo{volume}{65}},
  \bibinfo{pages}{851} (\bibinfo{year}{1976}).

\bibitem[{\citenamefont{Debenedetti}(2003)}]{De03}
\bibinfo{author}{\bibfnamefont{P.~G.} \bibnamefont{Debenedetti}},
  \bibinfo{journal}{J. Phys.: Cond. Matter} \textbf{\bibinfo{volume}{15}},
  \bibinfo{pages}{R1669} (\bibinfo{year}{2003}).

\bibitem[{\citenamefont{Scala et~al.}(2000)\citenamefont{Scala, Sadr-Lahijany,
  Giovambattista, Buldyrev, and Stanley}}]{Sc00}
\bibinfo{author}{\bibfnamefont{A.}~\bibnamefont{Scala}},
  \bibinfo{author}{\bibfnamefont{M.~R.} \bibnamefont{Sadr-Lahijany}},
  \bibinfo{author}{\bibfnamefont{N.}~\bibnamefont{Giovambattista}},
  \bibinfo{author}{\bibfnamefont{S.~V.} \bibnamefont{Buldyrev}},
  \bibnamefont{and} \bibinfo{author}{\bibfnamefont{H.~E.}
  \bibnamefont{Stanley}}, \bibinfo{journal}{J. Stat. Phys.}
  \textbf{\bibinfo{volume}{100}}, \bibinfo{pages}{97} (\bibinfo{year}{2000}).

\bibitem[{\citenamefont{Franzese et~al.}(2001)\citenamefont{Franzese, Malescio,
  Skibinsky, Buldyrev, and Stanley}}]{Fr01}
\bibinfo{author}{\bibfnamefont{G.}~\bibnamefont{Franzese}},
  \bibinfo{author}{\bibfnamefont{G.}~\bibnamefont{Malescio}},
  \bibinfo{author}{\bibfnamefont{A.}~\bibnamefont{Skibinsky}},
  \bibinfo{author}{\bibfnamefont{S.~V.} \bibnamefont{Buldyrev}},
  \bibnamefont{and} \bibinfo{author}{\bibfnamefont{H.~E.}
  \bibnamefont{Stanley}}, \bibinfo{journal}{Nature (London)}
  \textbf{\bibinfo{volume}{409}}, \bibinfo{pages}{692} (\bibinfo{year}{2001}).

\bibitem[{\citenamefont{Buldyrev et~al.}(2002)\citenamefont{Buldyrev, Franzese,
  Giovambattista, Malescio, Sadr-Lahijany, Scala, Skibinsky, and
  Stanley}}]{Bu02}
\bibinfo{author}{\bibfnamefont{S.~V.} \bibnamefont{Buldyrev}},
  \bibinfo{author}{\bibfnamefont{G.}~\bibnamefont{Franzese}},
  \bibinfo{author}{\bibfnamefont{N.}~\bibnamefont{Giovambattista}},
  \bibinfo{author}{\bibfnamefont{G.}~\bibnamefont{Malescio}},
  \bibinfo{author}{\bibfnamefont{M.~R.} \bibnamefont{Sadr-Lahijany}},
  \bibinfo{author}{\bibfnamefont{A.}~\bibnamefont{Scala}},
  \bibinfo{author}{\bibfnamefont{A.}~\bibnamefont{Skibinsky}},
  \bibnamefont{and} \bibinfo{author}{\bibfnamefont{H.~E.}
  \bibnamefont{Stanley}}, \bibinfo{journal}{Physica A}
  \textbf{\bibinfo{volume}{304}}, \bibinfo{pages}{23} (\bibinfo{year}{2002}).

\bibitem[{\citenamefont{Buldyrev and Stanley}(2003)}]{Bu03}
\bibinfo{author}{\bibfnamefont{S.~V.} \bibnamefont{Buldyrev}} \bibnamefont{and}
  \bibinfo{author}{\bibfnamefont{H.~E.} \bibnamefont{Stanley}},
  \bibinfo{journal}{Physica A} \textbf{\bibinfo{volume}{330}},
  \bibinfo{pages}{124} (\bibinfo{year}{2003}).

\bibitem[{\citenamefont{Skibinsky et~al.}(2005)\citenamefont{Skibinsky,
  Buldyrev, Franzese, Malescio, and Stanley}}]{Sk04}
\bibinfo{author}{\bibfnamefont{A.}~\bibnamefont{Skibinsky}},
  \bibinfo{author}{\bibfnamefont{S.~V.} \bibnamefont{Buldyrev}},
  \bibinfo{author}{\bibfnamefont{G.}~\bibnamefont{Franzese}},
  \bibinfo{author}{\bibfnamefont{G.}~\bibnamefont{Malescio}}, \bibnamefont{and}
  \bibinfo{author}{\bibfnamefont{H.~E.} \bibnamefont{Stanley}},
  \bibinfo{journal}{Phys. Rev. E} \textbf{\bibinfo{volume}{69}},
  \bibinfo{pages}{061206} (\bibinfo{year}{2005}).

\bibitem[{\citenamefont{Franzese et~al.}(2002)\citenamefont{Franzese, Malescio,
  Skibinsky, Buldyrev, and Stanley}}]{Fr02}
\bibinfo{author}{\bibfnamefont{G.}~\bibnamefont{Franzese}},
  \bibinfo{author}{\bibfnamefont{G.}~\bibnamefont{Malescio}},
  \bibinfo{author}{\bibfnamefont{A.}~\bibnamefont{Skibinsky}},
  \bibinfo{author}{\bibfnamefont{S.~V.} \bibnamefont{Buldyrev}},
  \bibnamefont{and} \bibinfo{author}{\bibfnamefont{H.~E.}
  \bibnamefont{Stanley}}, \bibinfo{journal}{Phys. Rev. E}
  \textbf{\bibinfo{volume}{66}}, \bibinfo{pages}{051206}
  (\bibinfo{year}{2002}).

\bibitem[{\citenamefont{Balladares and Barbosa}(2004)}]{Ba04}
\bibinfo{author}{\bibfnamefont{A.}~\bibnamefont{Balladares}} \bibnamefont{and}
  \bibinfo{author}{\bibfnamefont{M.~C.} \bibnamefont{Barbosa}},
  \bibinfo{journal}{J. Phys.: Cond. Matter} \textbf{\bibinfo{volume}{16}},
  \bibinfo{pages}{8811} (\bibinfo{year}{2004}).

\bibitem[{\citenamefont{de~Oliveira and Barbosa}(2005)}]{Ol05}
\bibinfo{author}{\bibfnamefont{A.~B.} \bibnamefont{de~Oliveira}}
  \bibnamefont{and} \bibinfo{author}{\bibfnamefont{M.~C.}
  \bibnamefont{Barbosa}}, \bibinfo{journal}{J. Phys.: Cond. Matter}
  \textbf{\bibinfo{volume}{17}}, \bibinfo{pages}{399} (\bibinfo{year}{2005}).

\bibitem[{\citenamefont{Henriques and Barbosa}(2005)}]{He05a}
\bibinfo{author}{\bibfnamefont{V.~B.} \bibnamefont{Henriques}}
  \bibnamefont{and} \bibinfo{author}{\bibfnamefont{M.~C.}
  \bibnamefont{Barbosa}}, \bibinfo{journal}{Phys. Rev. E}
  \textbf{\bibinfo{volume}{71}}, \bibinfo{pages}{031504}
  (\bibinfo{year}{2005}).

\bibitem[{\citenamefont{Henriques et~al.}(2005)\citenamefont{Henriques,
  Guissoni, Barbosa, Thielo, and Barbosa}}]{He05b}
\bibinfo{author}{\bibfnamefont{V.~B.} \bibnamefont{Henriques}},
  \bibinfo{author}{\bibfnamefont{N.}~\bibnamefont{Guissoni}},
  \bibinfo{author}{\bibfnamefont{M.~A.} \bibnamefont{Barbosa}},
  \bibinfo{author}{\bibfnamefont{M.}~\bibnamefont{Thielo}}, \bibnamefont{and}
  \bibinfo{author}{\bibfnamefont{M.~C.} \bibnamefont{Barbosa}},
  \bibinfo{journal}{Mol. Phys.} \textbf{\bibinfo{volume}{103}},
  \bibinfo{pages}{3001} (\bibinfo{year}{2005}).

\bibitem[{\citenamefont{Hemmer and Stell}(1970)}]{He70}
\bibinfo{author}{\bibfnamefont{P.~C.} \bibnamefont{Hemmer}} \bibnamefont{and}
  \bibinfo{author}{\bibfnamefont{G.}~\bibnamefont{Stell}},
  \bibinfo{journal}{Phys. Rev. Lett.} \textbf{\bibinfo{volume}{24}},
  \bibinfo{pages}{1284} (\bibinfo{year}{1970}).

\bibitem[{\citenamefont{Jagla}(1998)}]{Ja98}
\bibinfo{author}{\bibfnamefont{E.~A.} \bibnamefont{Jagla}},
  \bibinfo{journal}{Phys. Rev. E} \textbf{\bibinfo{volume}{58}},
  \bibinfo{pages}{1478} (\bibinfo{year}{1998}).

\bibitem[{\citenamefont{Wilding and Magee}(2002)}]{Wi02}
\bibinfo{author}{\bibfnamefont{N.~B.} \bibnamefont{Wilding}} \bibnamefont{and}
  \bibinfo{author}{\bibfnamefont{J.~E.} \bibnamefont{Magee}},
  \bibinfo{journal}{Phys. Rev. E} \textbf{\bibinfo{volume}{66}},
  \bibinfo{pages}{031509} (\bibinfo{year}{2002}).

\bibitem[{\citenamefont{Maruyama et~al.}(2004)\citenamefont{Maruyama,
  Wakabayashi, and Oguni}}]{Ma04}
\bibinfo{author}{\bibfnamefont{S.}~\bibnamefont{Maruyama}},
  \bibinfo{author}{\bibfnamefont{K.}~\bibnamefont{Wakabayashi}},
  \bibnamefont{and} \bibinfo{author}{\bibfnamefont{M.}~\bibnamefont{Oguni}},
  \bibinfo{journal}{Aip Conf. Proceedings} \textbf{\bibinfo{volume}{708}},
  \bibinfo{pages}{675} (\bibinfo{year}{2004}).

\bibitem[{\citenamefont{Kurita and Tanaka}(2004)}]{Ku04}
\bibinfo{author}{\bibfnamefont{R.}~\bibnamefont{Kurita}} \bibnamefont{and}
  \bibinfo{author}{\bibfnamefont{H.}~\bibnamefont{Tanaka}},
  \bibinfo{journal}{Science} \textbf{\bibinfo{volume}{206}},
  \bibinfo{pages}{845} (\bibinfo{year}{2004}).

\bibitem[{\citenamefont{Xu et~al.}(2005)\citenamefont{Xu, Kumar, Buldyrev,
  Chen, Poole, Sciortino, and Stanley}}]{Xu05}
\bibinfo{author}{\bibfnamefont{L.}~\bibnamefont{Xu}},
  \bibinfo{author}{\bibfnamefont{P.}~\bibnamefont{Kumar}},
  \bibinfo{author}{\bibfnamefont{S.~V.} \bibnamefont{Buldyrev}},
  \bibinfo{author}{\bibfnamefont{S.-H.} \bibnamefont{Chen}},
  \bibinfo{author}{\bibfnamefont{P.}~\bibnamefont{Poole}},
  \bibinfo{author}{\bibfnamefont{F.}~\bibnamefont{Sciortino}},
  \bibnamefont{and} \bibinfo{author}{\bibfnamefont{H.~E.}
  \bibnamefont{Stanley}}, \bibinfo{journal}{Proc. Natl. Acad. Sci. U.S.A.}
  \textbf{\bibinfo{volume}{102}}, \bibinfo{pages}{16558}
  (\bibinfo{year}{2005}).

\bibitem[{\citenamefont{de~Oliveira
  et~al.}(2006{\natexlab{a}})\citenamefont{de~Oliveira, Netz, Colla, and
  Barbosa}}]{Ol06a}
\bibinfo{author}{\bibfnamefont{A.~B.} \bibnamefont{de~Oliveira}},
  \bibinfo{author}{\bibfnamefont{P.~A.} \bibnamefont{Netz}},
  \bibinfo{author}{\bibfnamefont{T.}~\bibnamefont{Colla}}, \bibnamefont{and}
  \bibinfo{author}{\bibfnamefont{M.~C.} \bibnamefont{Barbosa}},
  \bibinfo{journal}{J. Chem. Phys.} \textbf{\bibinfo{volume}{124}},
  \bibinfo{pages}{084505} (\bibinfo{year}{2006}{\natexlab{a}}).

\bibitem[{\citenamefont{de~Oliveira
  et~al.}(2006{\natexlab{b}})\citenamefont{de~Oliveira, Netz, Colla, and
  Barbosa}}]{Ol06b}
\bibinfo{author}{\bibfnamefont{A.~B.} \bibnamefont{de~Oliveira}},
  \bibinfo{author}{\bibfnamefont{P.~A.} \bibnamefont{Netz}},
  \bibinfo{author}{\bibfnamefont{T.}~\bibnamefont{Colla}}, \bibnamefont{and}
  \bibinfo{author}{\bibfnamefont{M.~C.} \bibnamefont{Barbosa}},
  \bibinfo{journal}{J. Chem. Phys.} \textbf{\bibinfo{volume}{125}},
  \bibinfo{pages}{124503} (\bibinfo{year}{2006}{\natexlab{b}}).

\bibitem[{\citenamefont{de~Oliveira et~al.}(2007)\citenamefont{de~Oliveira,
  Barbosa, and Netz}}]{Ol07}
\bibinfo{author}{\bibfnamefont{A.~B.} \bibnamefont{de~Oliveira}},
  \bibinfo{author}{\bibfnamefont{M.~C.} \bibnamefont{Barbosa}},
  \bibnamefont{and} \bibinfo{author}{\bibfnamefont{P.~A.} \bibnamefont{Netz}},
  \bibinfo{journal}{Physica A} \textbf{\bibinfo{volume}{386}},
  \bibinfo{pages}{744} (\bibinfo{year}{2007}).

\bibitem[{\citenamefont{de~Oliveira
  et~al.}(2008{\natexlab{a}})\citenamefont{de~Oliveira, Netz, and
  Barbosa}}]{Ol08a}
\bibinfo{author}{\bibfnamefont{A.~B.} \bibnamefont{de~Oliveira}},
  \bibinfo{author}{\bibfnamefont{P.~A.} \bibnamefont{Netz}}, \bibnamefont{and}
  \bibinfo{author}{\bibfnamefont{M.~C.} \bibnamefont{Barbosa}},
  \bibinfo{journal}{Euro. Phys. J. B} \textbf{\bibinfo{volume}{64}},
  \bibinfo{pages}{48} (\bibinfo{year}{2008}{\natexlab{a}}).

\bibitem[{\citenamefont{de~Oliveira
  et~al.}(2008{\natexlab{b}})\citenamefont{de~Oliveira, Franzese, Netz, and
  Barbosa}}]{Ol08b}
\bibinfo{author}{\bibfnamefont{A.~B.} \bibnamefont{de~Oliveira}},
  \bibinfo{author}{\bibfnamefont{G.}~\bibnamefont{Franzese}},
  \bibinfo{author}{\bibfnamefont{P.~A.} \bibnamefont{Netz}}, \bibnamefont{and}
  \bibinfo{author}{\bibfnamefont{M.~C.} \bibnamefont{Barbosa}},
  \bibinfo{journal}{J. Chem. Phys.} \textbf{\bibinfo{volume}{128}},
  \bibinfo{pages}{064901} (\bibinfo{year}{2008}{\natexlab{b}}).

\bibitem[{\citenamefont{de~Oliveira et~al.}(2009)\citenamefont{de~Oliveira,
  Netz, and Barbosa}}]{Ol09}
\bibinfo{author}{\bibfnamefont{A.~B.} \bibnamefont{de~Oliveira}},
  \bibinfo{author}{\bibfnamefont{P.~E.} \bibnamefont{Netz}}, \bibnamefont{and}
  \bibinfo{author}{\bibfnamefont{M.~C.} \bibnamefont{Barbosa}},
  \bibinfo{journal}{Europhys. Lett.} \textbf{\bibinfo{volume}{85}},
  \bibinfo{pages}{36001} (\bibinfo{year}{2009}).

\bibitem[{\citenamefont{Krekelberg et~al.}(2008)\citenamefont{Krekelberg,
  Mittal, Ganesan, and Truskett}}]{Kr08}
\bibinfo{author}{\bibfnamefont{W.~P.} \bibnamefont{Krekelberg}},
  \bibinfo{author}{\bibfnamefont{J.}~\bibnamefont{Mittal}},
  \bibinfo{author}{\bibfnamefont{V.}~\bibnamefont{Ganesan}}, \bibnamefont{and}
  \bibinfo{author}{\bibfnamefont{T.~M.} \bibnamefont{Truskett}},
  \bibinfo{journal}{Phys. Rev. E} \textbf{\bibinfo{volume}{77}},
  \bibinfo{pages}{041201} (\bibinfo{year}{2008}).

\bibitem[{\citenamefont{Head-Gordon and Stillinger}(1993)}]{He93}
\bibinfo{author}{\bibfnamefont{T.}~\bibnamefont{Head-Gordon}} \bibnamefont{and}
  \bibinfo{author}{\bibfnamefont{F.~H.} \bibnamefont{Stillinger}},
  \bibinfo{journal}{J. Chem. Phys.} \textbf{\bibinfo{volume}{98}},
  \bibinfo{pages}{3313} (\bibinfo{year}{1993}).

\bibitem[{\citenamefont{Yan et~al.}(2008)\citenamefont{Yan, Buldyrev, Kumar,
  Giovambattista, and Stanley}}]{Ya08}
\bibinfo{author}{\bibfnamefont{Z.~Y.} \bibnamefont{Yan}},
  \bibinfo{author}{\bibfnamefont{S.~V.} \bibnamefont{Buldyrev}},
  \bibinfo{author}{\bibfnamefont{P.}~\bibnamefont{Kumar}},
  \bibinfo{author}{\bibfnamefont{N.}~\bibnamefont{Giovambattista}},
  \bibnamefont{and} \bibinfo{author}{\bibfnamefont{H.~E.}
  \bibnamefont{Stanley}}, \bibinfo{journal}{Phys. Rev. E}
  \textbf{\bibinfo{volume}{77}}, \bibinfo{pages}{042201}
  (\bibinfo{year}{2008}).

\bibitem[{\citenamefont{Stanley et~al.}(1998)\citenamefont{Stanley, Buldyrev,
  Canpolat, Meyer, Mishima, Sadr-Lahijany, Scala, and Starr}}]{St98}
\bibinfo{author}{\bibfnamefont{H.~E.} \bibnamefont{Stanley}},
  \bibinfo{author}{\bibfnamefont{S.~V.} \bibnamefont{Buldyrev}},
  \bibinfo{author}{\bibfnamefont{M.}~\bibnamefont{Canpolat}},
  \bibinfo{author}{\bibfnamefont{M.}~\bibnamefont{Meyer}},
  \bibinfo{author}{\bibfnamefont{O.}~\bibnamefont{Mishima}},
  \bibinfo{author}{\bibfnamefont{M.~R.} \bibnamefont{Sadr-Lahijany}},
  \bibinfo{author}{\bibfnamefont{A.}~\bibnamefont{Scala}}, \bibnamefont{and}
  \bibinfo{author}{\bibfnamefont{F.~W.} \bibnamefont{Starr}},
  \bibinfo{journal}{Physica A} \textbf{\bibinfo{volume}{257}},
  \bibinfo{pages}{213} (\bibinfo{year}{1998}).

\bibitem[{\citenamefont{Stanley}(1999)}]{St99}
\bibinfo{author}{\bibfnamefont{H.~E.} \bibnamefont{Stanley}},
  \bibinfo{journal}{[Proceedings of the 1998 International Conference on
  Complex Fluids], Pramana [A Journal of the Indian Academy of Sciences,
  founded by C. V. Raman]} \textbf{\bibinfo{volume}{53}}, \bibinfo{pages}{53}
  (\bibinfo{year}{1999}).

\bibitem[{\citenamefont{Stanley et~al.}(2000)\citenamefont{Stanley, Buldyrev,
  Canpolat, Mishima, Sadr-Lahijany, and Starr}}]{St00}
\bibinfo{author}{\bibfnamefont{H.~E.} \bibnamefont{Stanley}},
  \bibinfo{author}{\bibfnamefont{S.~V.} \bibnamefont{Buldyrev}},
  \bibinfo{author}{\bibfnamefont{M.}~\bibnamefont{Canpolat}},
  \bibinfo{author}{\bibfnamefont{O.}~\bibnamefont{Mishima}},
  \bibinfo{author}{\bibfnamefont{A.}~\bibnamefont{Sadr-Lahijany},
  \bibfnamefont{M.~R.~Scala}}, \bibnamefont{and}
  \bibinfo{author}{\bibfnamefont{F.~W.} \bibnamefont{Starr}},
  \bibinfo{journal}{Physical Chemistry and Chemical Physics}
  \textbf{\bibinfo{volume}{2}}, \bibinfo{pages}{1551} (\bibinfo{year}{2000}).

\bibitem[{\citenamefont{Errington et~al.}(2006)\citenamefont{Errington,
  Truskett, and Mittal}}]{Er06}
\bibinfo{author}{\bibfnamefont{J.~R.} \bibnamefont{Errington}},
  \bibinfo{author}{\bibfnamefont{T.~M.} \bibnamefont{Truskett}},
  \bibnamefont{and} \bibinfo{author}{\bibfnamefont{J.}~\bibnamefont{Mittal}},
  \bibinfo{journal}{J. Chem. Phys.} \textbf{\bibinfo{volume}{125}},
  \bibinfo{pages}{244502} (\bibinfo{year}{2006}).

\bibitem[{\citenamefont{Errington et~al.}(2003)\citenamefont{Errington,
  Debenedetti, and Torquato}}]{Er03}
\bibinfo{author}{\bibfnamefont{J.~E.} \bibnamefont{Errington}},
  \bibinfo{author}{\bibfnamefont{P.~G.} \bibnamefont{Debenedetti}},
  \bibnamefont{and} \bibinfo{author}{\bibfnamefont{S.}~\bibnamefont{Torquato}},
  \bibinfo{journal}{J. Chem. Phys.} \textbf{\bibinfo{volume}{118}},
  \bibinfo{pages}{2256} (\bibinfo{year}{2003}).

\bibitem[{\citenamefont{Camp}(2003)}]{Ca03}
\bibinfo{author}{\bibfnamefont{P.}~\bibnamefont{Camp}}, \bibinfo{journal}{Phys.
  Rev. E} \textbf{\bibinfo{volume}{68}}, \bibinfo{pages}{061506}
  (\bibinfo{year}{2003}).

\bibitem[{\citenamefont{Sadr-Lahijany et~al.}(1998)\citenamefont{Sadr-Lahijany,
  Scala, Buldyrev, and Stanley}}]{La98}
\bibinfo{author}{\bibfnamefont{M.~R.} \bibnamefont{Sadr-Lahijany}},
  \bibinfo{author}{\bibfnamefont{A.}~\bibnamefont{Scala}},
  \bibinfo{author}{\bibfnamefont{S.~V.} \bibnamefont{Buldyrev}},
  \bibnamefont{and} \bibinfo{author}{\bibfnamefont{H.~E.}
  \bibnamefont{Stanley}}, \bibinfo{journal}{Phys. Rev. Lett.}
  \textbf{\bibinfo{volume}{81}}, \bibinfo{pages}{4895} (\bibinfo{year}{1998}).

\end{thebibliography}

\end{document}